\documentclass[sigconf,screen]{acmart}

\let\xbibsection\bibsection
\renewcommand{\bibsection}{\xbibsection \emph{References marked with • are in the set of reviewed papers.}}

\usepackage{ifthen}
\usepackage{acronym}
\usepackage{tabularx}
\usepackage{subfig}
\usepackage{comment}
\usepackage{booktabs} 
\usepackage{ccicons}  
\usepackage[autostyle=true,english=american]{csquotes}
\usepackage{multirow}
\usepackage[english]{babel}

\graphicspath{{figures/}}

\newacro{AI}[AI]{Artificial Intelligence}
\newacro{UI}[UI]{user interface}
\newacro{GUI}[GUI]{graphical user interface}
\newacro{TLX}[TLX]{NASA-Task Load Index}
\newacro{RTLX}[Raw-TLX]{NASA Raw-Task Load Index}
\newacro{ER}[ER]{error rate}
\newacro{TCT}[TCT]{task completion time}
\newacro{HCI}[HCI]{Human-Computer Interaction}
\newacro{UX}[UX]{user experience}
\newacro{HFE}[HFE]{Human Factors and Ergonomics}
\newacro{cuDNN}[cuDNN]{CUDA Deep Neural Network library}
\newacro{RMSE}[RMSE]{root mean squared error}
\newacro{HMD}[HMD]{Head-Mounted Display}
\newacro{RF}[RF]{Random Forest}
\newacro{GP}[GP]{Gaussian process, long-plural = Gaussian processes}
\newacro{KNN}[\textit{k}NN]{\textit{k}-nearest neighbor}
\newacro{NN}[NN]{Neural Network}
\newacro{DNN}[DNN]{ Deep Neural Network}
\newacro{CNN}[CNN]{Convolutional Neural Network}
\newacro{FCL}[FCL]{fully connected layer}
\newacro{BoD}[BoD]{Back-of-Device}
\newacro{FOV}[FoV]{field of view}
\newacro{RW}[RW]{Real World}
\newacro{IFRC}[IFRC]{index finger ray cast}
\newacro{FRC}[FRC]{forearm ray cast}
\newacro{EFRC}[EFRC]{eye-finger ray cast}
\newacro{HRC}[HRC]{Human-Robot Collaboration}
\newacro{HRI}[HRI]{Human-Robot Interaction}
\newacro{6DOF}[6DOF]{six-degree-of-freedom}
\newacro{LOOCV}[LOOCV]{leave-one-out cross-validation}
\newacro{CV}[CV]{cross-validation}
\newacro{RM}[RM]{repeated measure}
\newacro{ANOVA}[ANOVA]{analysis of variance}
\newacro{RMANOVA}[RM-ANOVA]{repeated measures analysis of variance}
\newacro{AGATe}[AGATe]{AGreement Analysis Toolkit}
\newacro{GHoST}[GHoST]{Gesture Heatmap Toolkit Gesture Heatmaps Toolkit}
\newacro{GREAT}[GREAT]{Gesture Relative Accuracy Toolkit}
\newacro{GRT}[GRT]{Gesture Recognition Toolkit}
\newacro{DTW}[DTW]{Dynamic Time Warping}
\newacro{LHRD}[LHRD]{large high resolution display}
\newacro{GEQ}[GEQ]{Game Experience Questionnaire}
\newacro{SPGQ}[SPGQ]{Social Presence Gaming Questionnaire}
\newacro{JND}[JND]{just-noticeable difference}
\newacro{SUS}[SUS]{system usability scale}
\newacro{CSCW}[CSCW]{computer-supported cooperative work}
\newacro{CAD}[CAD]{computer-aided design}
\newacro{MR}[MR]{Mixed Reality}
\newacro{CVE}[CVE]{Collaborative Virtual Environment}
\newacro{AR}[AR]{Augmented Reality}
\newacro{AV}[AV]{Augmented Virtuality}
\newacro{VR}[VR]{Virtual Reality}
\newacro{PRISMA}[PRISMA]{Preferred Reporting Items for Systematic Reviews}
\newacro{PRISMA-Scope}[PRISMA-ScR]{Meta-Analyses Extension for Scoping Reviews}
\newacro{TF-IDF}[TF-IDF]{Term Frequency-Inverse Document Frequency}
\newacro{TF}[TF]{Term Frequency}
\newacro{AVs}[AVs]{Automated Vehicles}
\newacro{eHMIs}[eHMIs]{external Human-machine interfaces}
\newacro{SAR}[SAR]{Spatial Augmented Reality}
\newacro{IFR}[IFR]{International Federation of Robotics}
\newacro{ADLs}[ADLs]{Activities of Daily Living}
\newacro{LED}[LED]{Light-Emitting Diode}
\newacro{DoF}[DoF]{Degrees-of-Freedom}
\newacro{HHC}[HHC]{Human-Human Collaboration}
\newacro{IDF}[IDF]{Inverse Document Frequency}

\AtBeginDocument{%
  \providecommand\BibTeX{{%
    \normalfont B\kern-0.5em{\scshape i\kern-0.25em b}\kern-0.8em\TeX}}}

\copyrightyear{2023}
\acmYear{2023}
\setcopyright{rightsretained}
\acmConference[CHI '23]{Proceedings of the 2023 CHI Conference on Human Factors in Computing Systems}{April 23--28, 2023}{Hamburg, Germany}
\acmBooktitle{Proceedings of the 2023 CHI Conference on Human Factors in Computing Systems (CHI '23), April 23--28, 2023, Hamburg, Germany}
\acmDOI{10.1145/3544548.3580857}
\acmISBN{978-1-4503-9421-5/23/04}

\newcommand\change[1]{{#1}}
\begin{document}

\title{How to Communicate Robot Motion Intent: A Scoping Review}

\author{Max Pascher}
\orcid{0000-0002-6847-0696} 
\email{max.pascher@w-hs.de}
\affiliation{
    \institution{Westphalian University of Applied Sciences}
    \city{Gelsenkirchen}
    \country{Germany}
}
\affiliation{
    \institution{University of Duisburg-Essen}
    \city{Essen}
    \country{Germany}
}

\author{Uwe Gruenefeld}
\orcid{0000-0002-5671-1640}
\email{uwe.gruenefeld@uni-due.de}
\affiliation{
    \institution{University of Duisburg-Essen}
    \city{Essen}
    \country{Germany}
}

\author{Stefan Schneegass}
\orcid{0000-0002-0132-4934}
\email{stefan.schneegass@uni-due.de}
\affiliation{
    \institution{University of Duisburg-Essen}
    \city{Essen}
    \country{Germany}
}

\author{Jens Gerken}
\orcid{0000-0002-0634-3931}
\email{jens.gerken@w-hs.de}
\affiliation{
    \institution{Westphalian University of Applied Sciences}
    \city{Gelsenkirchen}
    \country{Germany}
}

\renewcommand{\shortauthors}{Max Pascher et al.}

\begin{abstract}
Robots are becoming increasingly omnipresent in our daily lives, supporting us and carrying out autonomous tasks.
In Human-Robot Interaction, human actors benefit from understanding the robot's motion intent to avoid task failures and foster collaboration.
Finding effective ways to communicate this intent to users has recently received increased research interest.
However, no common language has been established to systematize \emph{robot motion intent}.
This work presents a scoping review aimed at unifying existing knowledge. Based on our analysis, we present an intent communication model that depicts the relationship between robot and human through different intent dimensions (\emph{intent type}, \emph{intent information}, \emph{intent location}). We discuss these different intent dimensions and their interrelationships with different kinds of robots and human roles. Throughout our analysis, we classify the existing research literature along our intent communication model, allowing us to identify key patterns and possible directions for future research. 
\end{abstract}



\begin{teaserfigure}
  \subfloat[\emph{Motion}~\cite{Tsamis.2021}]{\includegraphics[width=0.245\linewidth]{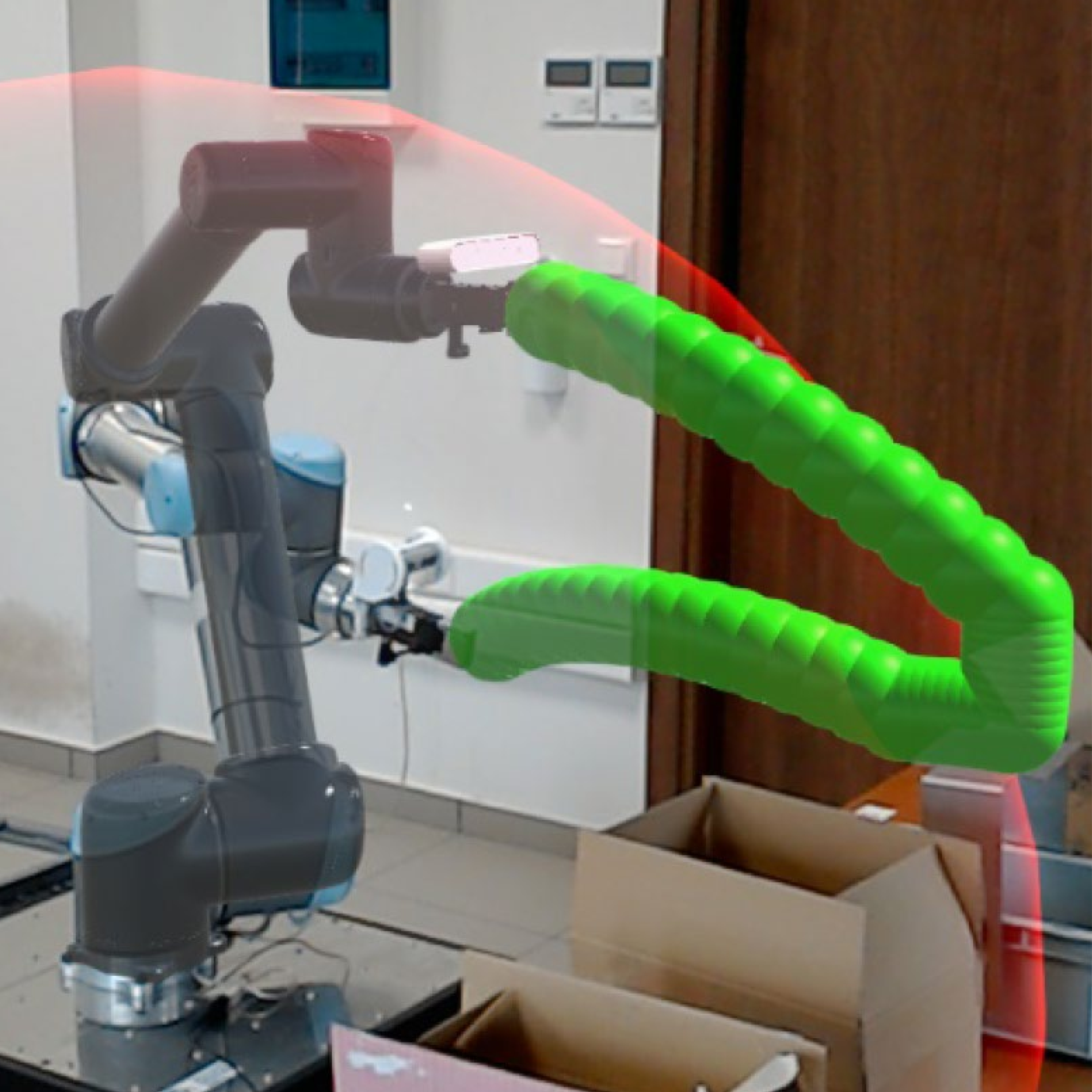}}
  \hfill
  \subfloat[\emph{Attention}~\cite{Koay.2013}]{\includegraphics[width=0.245\linewidth]{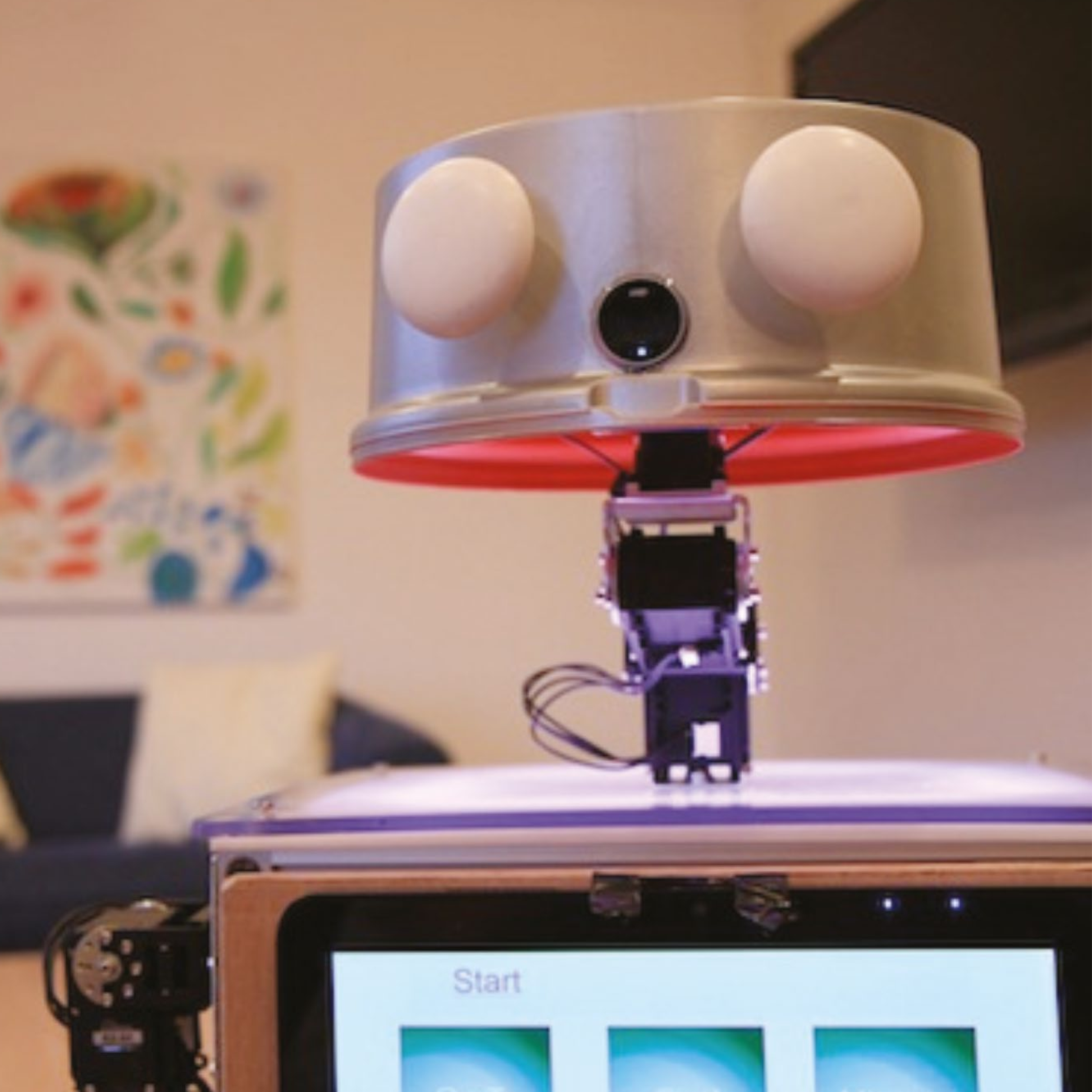}}
  \hfill
  \subfloat[\emph{State}~\cite{Tang.2019}]{\includegraphics[width=0.245\linewidth]{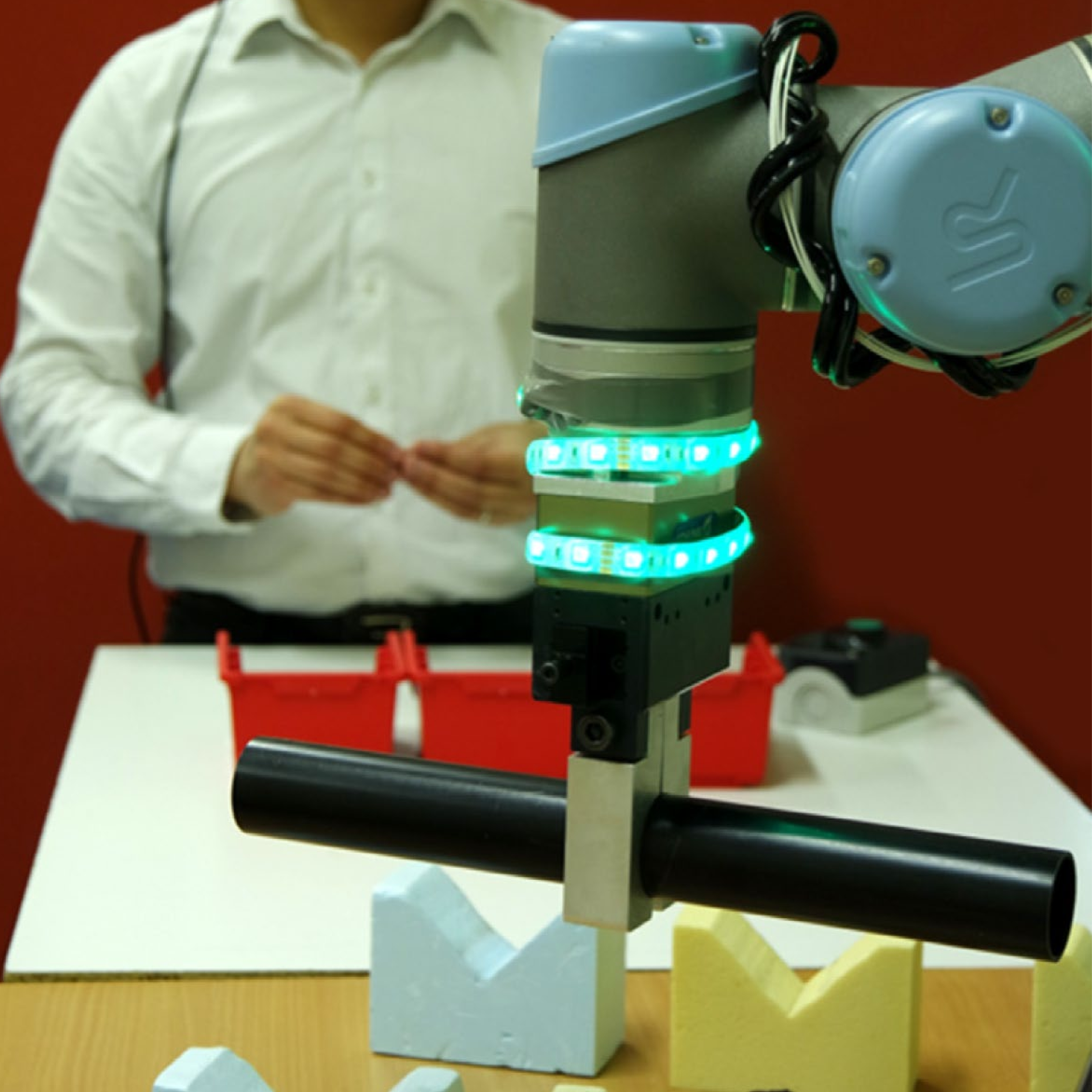}}
  \hfill
  \subfloat[\emph{Instruction}~\cite{Wengefeld.2020}]{\includegraphics[width=0.245\linewidth]{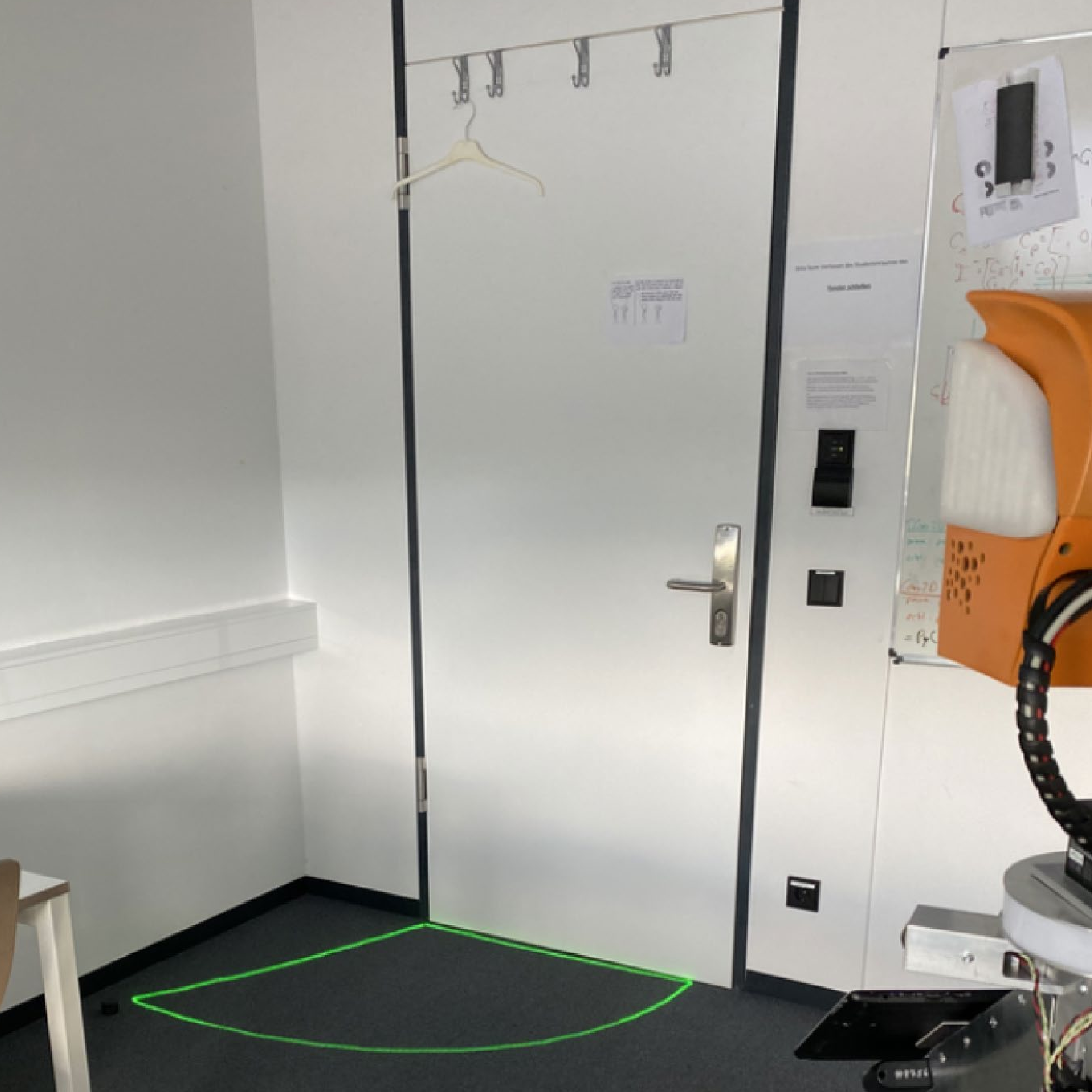}}
  \caption{(\textbf{a}) \emph{Robot motion intent}: The robot communicates its intended motion (e.g., a trajectory of the robot's intended movement path is visualized in Augmented Reality~\cite{Tsamis.2021}). Furthermore, our analysis revealed three additional types of intent that complement robot motion intents. (\textbf{b}) \emph{Attention}: A robot aims to catch the user's attention for subsequent movement activity (e.g., by moving its whole body~\cite{Koay.2013}). (\textbf{c}) \emph{State}: A robot communicates its state so that a human can predict future motions and identify potential conflicts before they occur (e.g., the robot communicates its movement activity with the help of a colored LED stripe~\cite{Tang.2019}). (\textbf{d}) \emph{Instruction}: The robot aims to provide specific instructions so that the human can assist further movement (e.g., by requesting to open a door~\cite{Wengefeld.2020}).} 
  \Description[Teaser figure of the paper]{An overview of identified robot motion intent types and corresponding examples from the literature. Figure 1a: a robotic arm mounted on a mobile platform. By Augmented Reality, another virtual robot is shown as an overlay. Illustrating the intended movement, a path of green spheres illustrates a planned trajectory, and a big semi-dense sphere is visualized around the arm, illustrating the maximum movement in each direction. Figure 1b: a human-like mobile robot with a head, two eyes, and no mouth mounted on a grey square-like body with a display attached to the front. The robot is looking into the camera. Figure 1c: a robotic arm is shown with a LED stripe wrapped around it. The LED stripe can light up in different colors to communicate different states. Figure 1d: an image of a room with a closed door. A schematic open door is shown via projection on the floor in front of the door.}
  \label{fig:teaser}
\end{teaserfigure}

\maketitle

\section{Introduction}
\label{sec:introduction}

The field of \ac{HCI} has moved beyond traditional user interfaces and interaction technologies. 
The omnipresence of \ac{AI} research and development requires our field to scrutinize the applicability of established design practices~\cite{amershi2019HumanAI,shneiderman2022human}. 
Human interaction with \ac{AI} is evolving away from being like operating a tool to being more like interacting with a partner, which is particularly interesting concerning \ac{HRI}~\cite{grudin2017tool}. 
\change{The area of \ac{HRI} has been studied for a long time in \ac{HCI} and, in particular, the CHI community~\cite{Noguchi2020,Arevalo.2021,Kim.2020,Liu.2011, Villanueva.2021}. 
For example, \citeauthor{Arevalo.2021}~\cite{Arevalo.2021} and \citeauthor{Villanueva.2021}~\cite{Villanueva.2021} investigated combining robots and \ac{AR} technology to enable intuitive teleoperation,
while others have explored on-site control of robot swarms~\cite{Kim.2020} and home robots~\cite{Liu.2011} as well as communication of emotions and intentions to the human~\cite{Noguchi2020}.

Robots are versatile, they} can assist us in our workplaces, support us at home, and accompany us in public spaces~\cite{Bauer.2008,Ajoudani.2017.collaboration,Mahdi.2022}. 
The applications of robots are manifold, significantly increasing human capabilities and efficiency~\cite{Galin.2020}. 
While robots come in many forms, robotic arms in particular have been shown to be suitable for and adaptable to different use cases, such as production lines~\cite{Braganca.2019} and domestic care~\cite{Pascher.2021}. 
Here, they are known as cobots who support their users in \ac{ADLs}, such as eating and drinking, grooming, or activities associated with leisure time. 

As robots have a physical form, they tend to move and operate in the same space as humans. With advances in the degree of autonomy allowing for effective close-contact interaction, there is a need for a shared understanding between humans and robots. 
While robotic research tackles this from a sensory and path planning perspective (e.g., human-aware navigation~\cite{KRUSE2013HumanAware}), \change{the field of \ac{HCI} (and \ac{HRI} in particular) has been concerned with how humans may better understand robot behavior~\cite{Rosen.2019,Walker.2018,Bodden.2018}}. 
The subtleties of human communication are usually lost in this context, and robotic behavior needs to be understood from its own frame of reference. 
Robots are not a monolithic entity; with the many different types come just as many unique ways of conveying information, which could lead to erroneous interpretations by their human counterpart.
An added complication is the increasing number of close-contact situations that allow little time to recognize and correct errors. 
This has led to numerous research efforts in recent years to find ways for robots to effectively communicate their intentions to their users~\cite{Kragic.2018}. 
This includes the direct communication of planned movements in space~\cite{Gruenefeld.2020}, but also less obvious means, such as drawing a user's attention to the robot~\cite{Koay.2013}, communicating the robot's movement activity state (e.g., active or inactive due to failure)~\cite{Song.2018}, and facilitating human oversight by communicating their external perception of the world~\cite{Han.2020}.

While all of these examples are concerned with communicating \emph{robot motion intent}, they differ tremendously in their methods and goals. 
Other researchers, such as \citeauthor{Suzuki.2022.AR-HRC-Survey}, have subsequently identified \emph{robot motion intent} as an essential research area~\cite{Suzuki.2022.AR-HRC-Survey}. 
But beyond further solution approaches, the field needs a common understanding of the concept of \emph{robot motion intent} (i.e., what do we actually mean by intent, what are relevant intent dimensions, and how does the communication of \emph{robot motion intent} influence the relationship between robot and human).

To this end, we conducted a scoping review of current approaches to communicate \emph{robot motion intent} in the literature. 
Based on our findings,  we introduce an intent communication model of \emph{motion intent}, which depicts the relationship between robot and human through the means of different intent dimensions (\emph{intent type}, \emph{intent information}, and \emph{intent location}; see~\autoref{fig:teaser}).
We further discuss these different intent dimensions and their interrelationships with different kinds of robots and human roles. 
Throughout our analysis, we classify the existing research literature along our intent communication model to form a design space for communicating \emph{robot motion intent}. 
Practitioners and researchers alike may further benefit from this work for the design and selection of specific mechanisms to communicate \emph{motion intent}. 
We identify future research directions and current gaps, which are further highlighted in an interactive website that lists the papers and allows comparisons based on user-selected categories.\footnote{\change{Interactive Data Visualization of the Paper Corpus~\cite{toby}. \url{https://rmi.robot-research.de}, last retrieved \today.}} 

\textbf{Our contribution} is two-fold: 1) a survey contribution that includes our analysis and classification of previous literature as well as future research (cf. contribution from~\citeauthor{Wobbrock.2016contributionsInHCI}~\cite{Wobbrock.2016contributionsInHCI}), and 2) a theoretical contribution that introduces an intent communication model and describes the relationship of its entities. 
\section{Background}
\label{sec:background}
In this section, we will illustrate the need for communicating \emph{robot motion intent} and discuss the current understanding of the term, which provides the foundation for our scoping review.

\emph{Robot} is an umbrella term that describes a miscellaneous collection of (semi-)automated devices with various capabilities, technologies, and appearances\cite{Goodrich.2008}. 
These cyber-physical systems are often differentiated by their \ac{DoF} or ability to move and manipulate their environment. 
In industrial assembly lines, robotic arms manipulate and weld heavy parts~\cite{Wang.2019}, often in restricted areas~\cite{Hentout.2019.Collaboration}. 
Enabled by lightweight materials and safety sensors, robots have started to adapt to their users -- today, they shut down when humans get too close or when resistance to the robot's movement is detected.
This has led to the development of \emph{cell-less} \ac{HRI}~\cite{Bauer.2016.Collaboration}, which has also paved the way for further scenarios, such as supporting people with disabilities in their daily lives~\cite{Pascher.2019b}. 
\citeauthor{Ajoudani.2017.collaboration} trace in their review paper several approaches of \ac{HRI}, how it evolved, and how it increased over the last two decades~\cite{Ajoudani.2017.collaboration}. 
They conclude that the success of \ac{HRI} comes from combining human cognitive skills (i.e., intelligence, flexibility, and ability to act in case of unexpected events) with the robot's high precision and ability to perform repetitive tasks.

\citeauthor{Matheson.2019.Collaboration} proposed different types of such \emph{cell-less} \ac{HRI}, defined by their closeness of interaction~\cite{Matheson.2019.Collaboration}. 
They include \emph{coexistence} (separation in space but not in time), \emph{synchronized} (no separation in space but in time), \emph{cooperation} (no separation in space or in time, but still not working on the same task), and \emph{collaboration} (human and robot work on a task together, where the action of one has immediate consequences for the other). 
These works indicate that communication and interaction between robots and humans are critical to successful \ac{HRI}. 
While research in human-aware navigation aims to make the robot smart enough to understand human behavior and react to it~\cite{KRUSE2013HumanAware}, supporting humans in understanding robot behavior is equally important~\cite{Kragic.2018}. 
As the work by \citeauthor{Matheson.2019.Collaboration} highlights, humans and robots increasingly share the same physical space in \ac{HRI}, which makes communicating \emph{robot motion intent} a particularly relevant aspect for safe and effective collaboration and a prerequisite for \emph{explainable robotics}~\cite{Matheson.2019.Collaboration}.

However, \emph{robot motion intent} is a rather vague term and lacks a clear definition. 
Further, it is not consistently used by researchers in the field. 
Instead, similar underlying concepts have been investigated under terms such as situational awareness~\cite{Levillain.2019}, forthcoming operation~\cite{Matsumaru.2007}, or robot signaling system~\cite{Tang.2019}.
\citeauthor{Suzuki.2022.AR-HRC-Survey}, as part of their extensive literature review covering the relationship between \ac{AR} and robotics, emphasize the potential of \ac{AR}-based visualizations for communicating movement trajectories or the internal state of the robot~\cite{Suzuki.2022.AR-HRC-Survey}. 
However, as their literature review extends beyond intent communication, they do not further discuss or define different types of intent, nor do they provide a deeper understanding of intent properties.

\textbf{Our work} presents a systematic overview of the field and addresses the current issues by conducting a scoping review.
Such a review or survey contribution helps to organize the published research of the field and enables reflection on previous findings after the field has reached a level of maturity~\cite{Wobbrock.2016contributionsInHCI}. 
The goal of our review is to provide a clear understanding and definition of \emph{robot motion intent}, its properties, and its relationships within \ac{HRI}.
Furthermore, our work provides a first discussion to relate our \ac{HRI} findings to the growing domain of \ac{AVs}, so-called \ac{eHMIs}, which have identified similar research and design challenges~\cite{Bazilinskyy.2019.AutomotiveSurvey,Dey.2020.AutomotiveSurvey,Rouchitsas.2019.AutomotiveSurvey,Colley2021.AV.3AC,Currano2021.AV.3AC}. 
\section{Method}
\label{sec:method}
Scoping reviews provide an overview of the extent, range, and nature of evolving research areas. They help to summarize research findings and identify research opportunities \cite{VonElm2019, Arksey2005}. Our approach is in line with previous work by Ghafurian et al.~\cite{Ghafurian.2021}, Mu\~{n}oz et al.~\cite{Munoz.2021}, and Wallk\"{o}tter et al.~\cite{Wallkoetter.2021}. We applied \emph{\ac{PRISMA}}~\cite{Page.2021.prisma} guidelines, focusing on the \emph{\ac{PRISMA-Scope}}~\cite{Tricco.2018prisma-scr}. 

\begin{figure*}
    \centering
    \includegraphics[width=\linewidth]{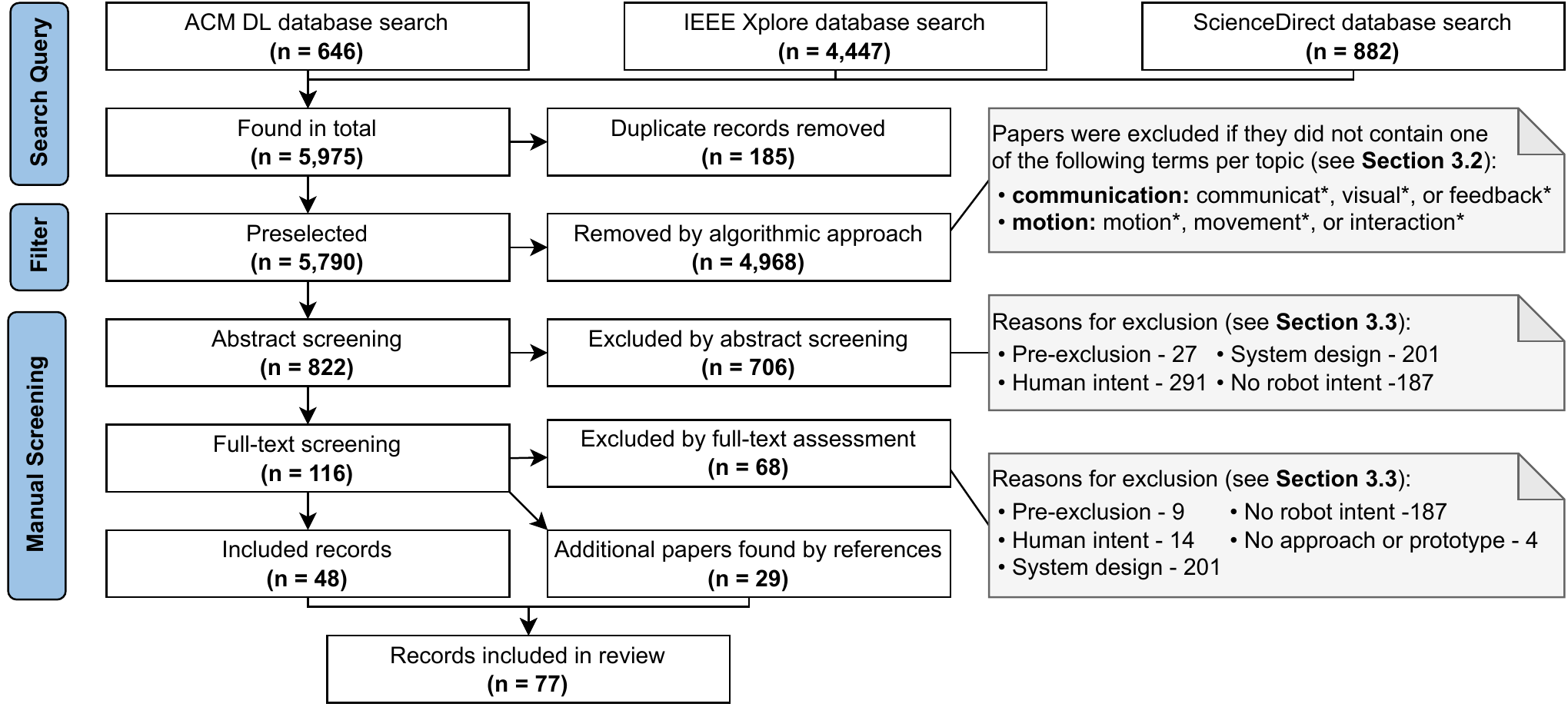}
    \caption{Flow chart of the corpus selection process with the identification of publishers and the initial search query (see~\autoref{sec:identificationOfSources}), the reduction of the set by algorithmic filtering (see~\autoref{sec:algorithmicApproach}), and the manual screening (see~\autoref{sec:screening}), which resulted in 77 papers.} 
    \Description[A flow chart illustrating the whole paper selection process to build the corpus]{From top to bottom, it is structured in 3 sections: Initial Search, Filtering, and Manual Screening. Initial Search: consisting of Papers found in ACM Digital Library (646), Papers found in IEEE Xplore (4447), and Papers found in ScienceDirect (882), resulting in 5975 papers and 185 duplicates were removed. Filtering: consisting of the number of preselected papers (5975 - 185) = 5790 minus 4968 papers excluded by the algorithmic approach. Manual Screening: This leads to 822 papers for the abstract screening. Other 706 papers were removed based on the criteria in Section 3.3. 116 Papers remain for the full-text screening. Another 68 Papers were removed based on the criteria in Section 3.3. This leads to 48 papers. From the full-text screening (116), another 29 papers were identified based on the reference screening, combined with the included 48 papers of the full-text screening leading to 77 papers in the corpus.}
    \label{fig:method:flowchart}
\end{figure*}

For an overview of each step in our paper selection process, please refer to~\autoref{fig:method:flowchart}. We will discuss specific details of the individual steps in the following subsections. 
(1) Based on an initial screening of relevant literature, potential search terms were identified to perform a systematic query using three primary databases in the field of \ac{HRI} (ACM Digital Library, IEEE Xplore, and ScienceDirect; see~\autoref{sec:identificationOfSources}).
(2) A filtering step was applied based on an algorithmic analysis of the total corpus to identify the most relevant terms related to the topic (see~\autoref{sec:algorithmicApproach}).
(3) The resulting set of 822 papers was manually screened in a two-step process, and eventually, additional sources were found through a cross-check of the references in selected papers (see~\autoref{sec:screening}). The final corpus consists of 77 papers. 

\subsection{Initial Query}
\label{sec:identificationOfSources}
We explored a variety of query terms and their combinations because, as discussed, the field currently lacks a coherent and established terminology. In addition, we found several terms to be used in ambiguous ways, in particular terms such as \emph{communication} and \emph{motion}. Therefore, we decided on a broad search in this first step to increase recall and reduce the risk of overlooking relevant literature. We aimed to encompass a variety of different robot technologies while still focusing on the concept of intent, even though the word may be used in a variety of circumstances. We searched the titles, abstracts, and keywords of the databases’ full-text collections with the following combined terms:\footnote{ScienceDirect does not support the wildcard \enquote{*} but uses stemming and lemmatization techniques. In order to achieve search results based on wildcards \enquote{*,} we modified the combined term to: \textit{(robot \textbf{OR} cobot \textbf{OR} drone) \textbf{AND} (intent \textbf{OR} intention \textbf{OR} intend \textbf{OR} intended)}.} 
\begin{eqnarray}
    \textit{(robot* \textbf{OR} cobot* \textbf{OR} drone*) \textbf{AND} (intent* \textbf{OR} intend*)}
\end{eqnarray}

\subsection{Algorithmic Filtering} 
\label{sec:algorithmicApproach}
Due to our initial search being quite broad, further filtering was required to identify relevant papers. The initial set allowed us to apply an algorithmic approach similar to that of previous research done by \citeauthor{OMaraEves.2015.textmining}~\cite{OMaraEves.2015.textmining}. Specifically, we applied the \ac{TF-IDF}~\cite{Salton1988} method to identify frequently used terminology within our corpus. \ac{TF-IDF} has been shown to be suitable for information retrieval in literature reviews~\cite{Surian2021,Lerner2019}. First, we preprocessed the entries by a) combining each paper's title, keywords, and abstract into one field, b) fixing encoding issues such as \emph{\&} (and), \emph{°} (degree), and \emph{---} (emdash), and c) converting the strings to lowercase as well as removing punctuation, numbers, symbols, and standard English stop-words from the corpus and replacing tokens with their lemmatizations~\cite{Manning2008}. For the creation of the \ac{TF-IDF}-weighted document-term matrix, we calculated the \ac{TF} for each term of our corpus, taking the static \ac{IDF} into account, and computed the \ac{TF-IDF} for each term over all documents. The resulting \ac{TF-IDF}-weighted document-term matrix is shown in \autoref{tab:TF-IDF}.

\begin{table*}[htbp]
    \centering 
    \caption{Sorted list of terms from the \ac{TF-IDF}-weighted document-term matrix. The selected terms are highlighted in bold.} 
    \begin{tabular}{ccccc||ccccc}
    \toprule
    \textbf{Rank} & \textbf{Term} & \textbf{TF} & \textbf{IDF} & \textbf{TF-IDF} & \textbf{Rank} & \textbf{Term} & \textbf{TF} & \textbf{IDF} & \textbf{TF-IDF}\\
    \midrule
    1 & human & 6,547 & 0.92 & 6,052.89 & 7 & \textbf{interaction} & 3,383 & 1.33 & 4,515.61 \\
    2 & control & 6,769 & 0.87 & 5,902.24 & 15 & \textbf{movement} & 1,920 & 1.88 & 3,606.34 \\
    3 & system & 7,612 & 0.69 & 5,218.61 & 61 & \textbf{communicat} & 1,059 & 2.32 & 2,455.03 \\
    4 & \textbf{motion} & 3,640 & 1.42 & 5,154.59 & 140 & \textbf{feedback} & 665 & 2.74 & 1,820.08 \\
    5 & model & 3,978 & 1.24 & 4,938.74 & 143 & \textbf{visual} & 674 & 2.67 & 1,802.90 \\
    \bottomrule
    \end{tabular}
    \Description{Sorted list of terms from the TF-IDF-weighted document-term matrix. The terms "motion", "interaction", "movement", "communicat", "feedback", and "visual" are highlighted in bold.}
    \label{tab:TF-IDF}
\end{table*}

\noindent From the first 150 entries of the \ac{TF-IDF} sorted list of tokens, 
three researchers independently qualified related terms to \emph{communication} and \emph{motion} -- two terms we had decided to leave out of the initial broad query due to word ambiguity. During the following consensus process, we excluded related terms that were too general and ambiguous (e.g., \enquote{show} is frequently used in \enquote{Our results show[...],} \enquote{present} in \enquote{In this work we present[...],} \enquote{demonstrate} in \enquote{We demonstrate in our results[...],} or \enquote{perform} in \enquote{We performed a study[...]}). {All identified terms were then used in the filtering step by applying the following logic to the title, keywords, or abstract of each paper in our corpus:
\begin{eqnarray}
&&    \textit{(communicat* \textbf{OR} visual* \textbf{OR} feedback*)\nonumber}\\ 
&&\textit{\textbf{AND} (motion* \textbf{OR} movement* \textbf{OR} interaction*)}
\end{eqnarray}}
For a paper to be accepted, a term from the cluster \enquote{communication} and another from \enquote{motion} (both OR operation) had to appear in the title, keywords, or abstract (AND operation). As a result, 822 papers remained in our corpus.

\subsection{Manual Screening}
\label{sec:screening}
The final phase of our paper selection process required manual screening, following an approach similar to that of \citeauthor{doherty2018survey}~\cite{doherty2018survey}. The process involved \emph{abstract screening}, \emph{full-text screening}, and \emph{reference screening}. During the screening of all abstracts, we identified 706 out of 822 papers as not fitting into the scope of this review. The full-text analysis of the remaining 116 papers reduced the set to 48 papers. In addition, we screened the references cited by the set of 116 papers that were assessed for full-text screening. We identified 29 further relevant references, which we then included. This led to a final set of 77 papers, which were examined in the following. During the abstract and full-text screening, we \textbf{pre-excluded} 36 papers in unfitting paper formats still in the corpus, such as proceedings front matter, workshop calls, survey papers, or semi-duplicates -- when two papers essentially presented the same contribution, due to one being a work in progress and the other a full paper. We also excluded 305 papers that aimed to convey the \textbf{human's intent} (to the robot) but not the robot's intent (e.g., \citeauthor{Kurylo.2019}~\cite{Kurylo.2019}). Similarly, we removed another 210 papers where the research did not focus on the intention of robot motion (\textbf{no robot intent}). For example, 1:1 teleoperated devices (e.g., \citeauthor{Waveren.2019}~\cite{Waveren.2019}), or work focusing on \ac{AVs} and \ac{eHMIs}. We excluded another 220 \textbf{system design} papers that focused on aspectus such as aesthetics, mathematical models of motion planning, or definitions (e.g., \citeauthor{Girard.2015}~\cite{Girard.2015}). Eventually, we removed four papers where no approach or prototype was developed and reported (e.g., \citeauthor{Thellman.2021}~\cite{Thellman.2021}).
\section{Intent Communication Model} 
\label{sec:intentmodel}
\begin{figure*}
    \centering
    \includegraphics[width=\linewidth]{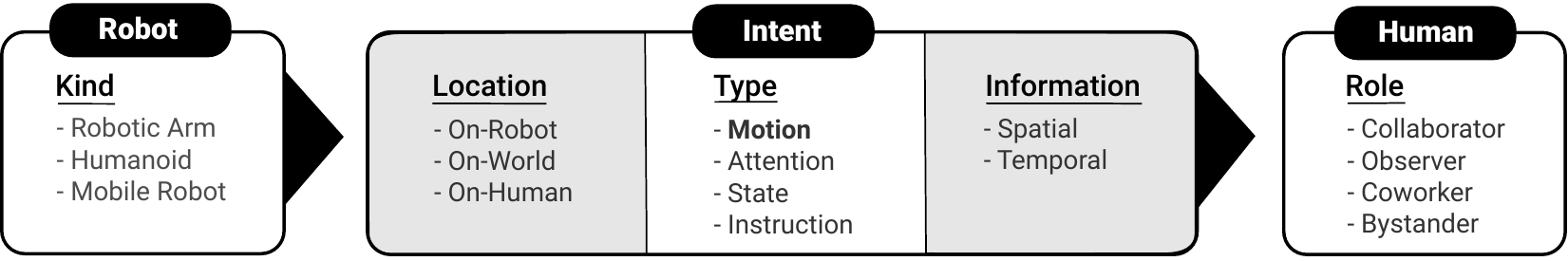}
    \caption{Overview of the intent communication model from robot to human. The three entities (i.e., robot, intent, human) and their dimensions are derived from our literature corpus. The flow of communication parallels the human-computer interaction model from Schomaker~\cite{schomaker1995taxonomy}. The main dimensions (i.e., kind, type, role) are discussed in \autoref{sec:intentmodel}, while a focused analysis of intent information and location is presented in \autoref{sec:indepthanalysis}.}
    \Description[Overview of our Intent Communication Flow Mode from robot to human]{Three horizontal main boxes describing the flow from left to right ("robot" -> "intent" -> "human"). Robot: Consisting of the kind of robot ("robotic arm", "humanoid", and "mobile robot"). Intent: Consisting of three further dimensions ("location", "type", "information"). Location including "On-Robot", "On-World", and "On-Human". Type including "Motion", "Attention", "State", and "Instruction". Information including "Spatial" and "Temporal". Human: Consisting of the roles of human ("Collaborator", "Observer", "Coworker", and "Bystander").}
    \label{fig:intentmodel}
\end{figure*}
Through our literature review, we aim to improve understanding of the communication of \emph{robot motion intent} by analyzing previous research.
To that end, each author analyzed our literature corpus (n=77) in a multi-step process. 
It was discovered that several papers presented, combined, or empirically compared multiple intents (on average, more than two per paper). Therefore, we first systematically extracted all individual intents, resulting in a total of 172 intents.
By screening these intents, we identified the primary entities (\emph{robot}, \emph{intent}, and \emph{human}) as well as a communication flow between these entities that parallels that of the \ac{HCI} model from \citeauthor{schomaker1995taxonomy}~\cite{schomaker1995taxonomy}. 
However, in contrast to the HCI model, we focus solely on the communication of \emph{intent} from \emph{robot} to \emph{human}, as previous research has already covered the inverse~\cite{Siddarth.2019}. 
Furthermore, we identified a top-level entity, \emph{goal}, which describes the motivation to communicate intent, as well as a low-level entity, \emph{context}, which describes the situation in which the intent is communicated.
Reflecting on all entities, we analyzed the intents by asking 1) \emph{why} they were communicated (\emph{goal}), 2) \emph{who} communicated them (\emph{robot}), 3) \emph{what} they communicated (\emph{intent}), 4) \emph{to whom} they were communicated (\emph{human}), and 5) \emph{in which} circumstances they were communicated (\emph{context}).
Dimensions, categories, and properties emerged from the data through an open coding process of the extracted answers; specifically, we identified \emph{kind of robot}, \emph{location}, \emph{type of intent}, \emph{information} of \emph{intent}, and \emph{role of human} as our dimensions. 
The resulting \emph{intent communication model} is shown in \autoref{fig:intentmodel}. 
In the following, we present our findings for the three primary entities (\emph{robot}, \emph{intent}, and \emph{human}), which we define and support by giving examples. We also discuss the \emph{context} of communicating \emph{robot motion intent}.

\subsection{Human}
\label{sub:intentmodel:human}
\begin{figure*}
    \centering
    \includegraphics[width=\linewidth]{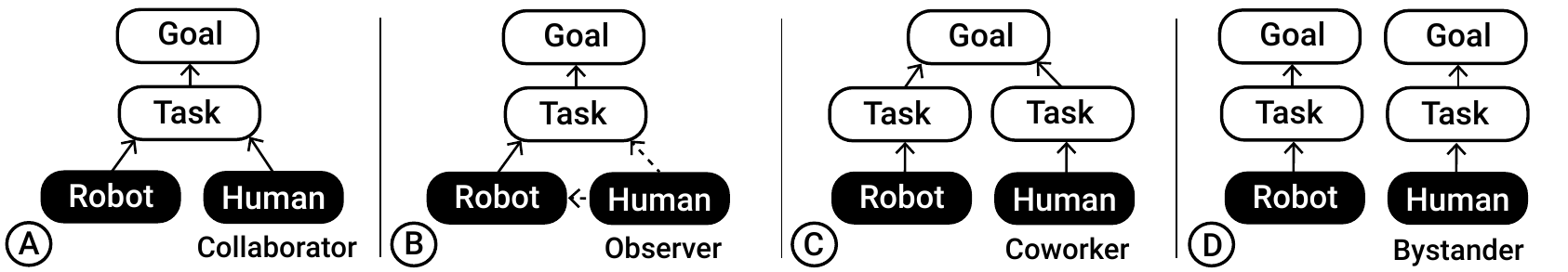}
    \caption{Comparison of the four human roles. The goals are further broken down by tasks to illustrate the relationships between the three entities (human, robot, and goals). A) The human (collaborator) and robot work on the same task. B) The human (observer) observes the robot and task but does not directly contribute. C) The human (coworker) and robot work on different tasks that contribute to the same goal. D) The human (bystander) and robot work on different tasks that contribute to different goals.}
    \Description[Comparison of the four human roles]{Four sub-figures illustrating the relationship between "robot", "human", "task", and the "overlying goal". Figure 4a: Collaborator: Robot and human are working together on the same task, which leads to the same goal. Figure 4b: Observer: Robot is working on a specific task. The human observes the robot and the task, following the same goal. Figure 4c: Coworker: Robot and human are working on separate tasks (alone) but both tasks are important for the same goal. Figure 4d: Bystander: Robot and human are working on separate tasks (alone) and following their own goals.}
    \label{fig:humanroles}
\end{figure*}

In \ac{HRI}, we can distinguish between different scenarios based on how involved a human is in the task performed by the robot. For the entity \emph{human}, we utilize these levels of closeness between robot and human to define the different \emph{roles of human}. 
Moreover, all four \emph{roles of human} are illustrated in~\autoref{fig:humanroles}.

\subsubsection{Definition}
\label{sub:intentmodel:human:definition}
The human has a crucial role during \ac{HRI}, which strongly impacts which intents need to be communicated. 
From the analyzed intents of our corpus, we derived four different \emph{roles of human} (\emph{collaborator}, \emph{observer}, \emph{coworker}, and \emph{bystander}). 
The roles are ordered by the degree of human collaboration and involvement with the robot, starting with the most involved~(see~\autoref{fig:humanroles}). These roles are also closely connected to the overarching goal of the \ac{HRI}. Here, we found \emph{supporting collaboration}, \emph{oversight}, and \emph{coexistence} to be of primary importance. In the following, we define the different roles, discuss their relationships to overarching goals, and support them with examples.

\paragraph{Collaborator}
\label{par:intentmodel:human:definition:collaborator}
When in the role of a \emph{collaborator}, a human works with a robot on a shared task in the same space and at the same time~\cite{Matheson.2019.Collaboration}. Thus, communication of \emph{robot motion intent} in this context is for \emph{supporting collaboration}. It aims to foster the coordination of robot and human actions regarding space and time to allow them to work together on a shared task (e.g., a human-robot assembly team in a manufacturing scenario~\cite{Andersen.2016}). The action of one of the two (i.e., robot or human) has immediate consequences for the other. For example, consider the scenario of a robot handing an object to a human~\cite{Newbury.2021, Dragan.2015}. Here, the human has to precisely anticipate and coordinate with the time and place the object will be positioned to enable efficient handover. To that end, \citeauthor{Dragan.2015} propose a robotic arm that applies so-called \emph{legible motion}, allowing the human to infer the goal of motion quickly and with certainty~\cite{Dragan.2015}. The role of a \emph{collaborator} represents the closest degree of \ac{HRI}, as they form a team in which both depend on each other. 
In our literature corpus, a \emph{collaborator} is described in 18 papers and is the recipient of 37 different intents.

\paragraph{Observer}
\label{par:intentmodel:human:definition:observer}
A human functions as an \emph{observer} when their main job is to supervise the task that is being carried out by the robot. Although they mostly just watch, an \emph{observer} must be ready to intervene and take control of the robot. 
In this context, communication of \emph{robot motion intent} is for the goal of \emph{supporting oversight}. Here, the robot has to provide information to the human to allow effective intervention when needed. Fundamentally, supporting oversight refers to the ability of a human to judge and evaluate if a robot is operating within its intended parameters. For example, in work by~\citeauthor{Hetherington.2021}, the robot communicates its movement paths to an \emph{observer}, which enables the \emph{observer} to foresee and prevent potential collisions of the robot with obstacles~\cite{Hetherington.2021}. Others communicate the inner state of the robot, allowing an \emph{observer} to anticipate potential task failures that may occur due to problems with the robot itself, e.g., faulty sensor information~\cite{Baraka.2016,Han.2020}. 
An \emph{observer} is described in 47 papers and is the recipient of 94 intents.

\paragraph{Coworker}
\label{par:intentmodel:human:definition:coworker}
In the \emph{coworker} role, the human works next to the robot but handles their own task. While these tasks may be part of a shared overarching effort or entirely disconnected, they take place in the same shared workspace (e.g., a robotic arm that picks up one out of two objects and leaves the other one for the human~\cite{LeeChang.2018}). In the \emph{coworker} context, communication of \emph{robot motion intent} is for the purpose of \emph{supporting coexistence}. Here, the human needs to understand the robot's motion to avoid safety-critical situations (e.g., colliding with the robot). In~\citeauthor{Aubert.2018}, a robot and human pick up objects from a shared bin for their individual tasks~\cite{Aubert.2018}. Here, communication of \emph{robot motion intent} can help the human to coordinate their actions and avoid collisions with the robot. \citeauthor{Chadalavada.2020} showed that communication of \emph{motion intent} through \ac{SAR} can improve perceived safety with mobile robots~\cite{Chadalavada.2020}. In their study, it meant that participants could choose safer walking paths and get closer to the robot without subsequent safety shutdowns. In our literature corpus, a \emph{coworker} is described in six papers and is the recipient of 18 intents.

\paragraph{Bystander}
\label{par:intentmodel:human:definition:bystander}
The human is a \emph{bystander} when they do not share the same task or the same task goal with the robot but still occupy an area overlapping the robot's physical workspace. Like the \emph{coworker} role, the \emph{bystander} role involves communication of \emph{robot motion intent} to support the goal of \emph{supporting coexistence}. A \emph{bystander} needs motion information to avoid collision and feel safe. For example, imagine a human and a robot encountering each other in a corridor. To allow the human to choose a walking path that avoids collision, the robot can move to one side and communicate its intended movement path in advance~\cite{Mikawa.2018, Watanabe.2015}. A \emph{bystander} is described in 17 papers and is the recipient of 23 intents.

\subsection{Intent}
\label{sub:intentmodel:intent}
We identified four different types of \emph{intent} that the \emph{robot} can communicate to the \emph{human} to express its intentions, contributing to increased transparency. We consider these types to be the main dimension for classifying \emph{intent} in the following text. In addition, we identified the dimensions \emph{location} 
and \emph{information}, 
as shown in~\autoref{fig:intentmodel}, which help to further classify and describe \emph{intent}. Given their great importance, they are discussed separately in~\autoref{sec:indepthanalysis}.

\subsubsection{Definition}
\label{sub:intentmodel:intent:definition}
As our literature review focused on communicating \emph{robot motion intent}, a majority of the corpus (69\% of all papers; 54\% of all unique intents) deals with \emph{motion intent}.
Nevertheless, we identified additional intent types that are related to \emph{motion intent} and of equal importance (i.e., \emph{attention}, \emph{state}, and \emph{instruction}).
All \emph{types of intent} are described below and the relationship of each to motion is explained.
Furthermore, we found that for each \emph{type of intent}, we can further distinguish between an \emph{intent} that is \emph{related to the robot} and one that is \emph{related to the world} (more details can be found in the individual paragraphs below).
An overview of all \emph{types of intent} and associated papers can be found in~\autoref{tab:intenttypes}.

\begin{table*} 
    \centering 
    \caption{Overview of different intent types, sorted by their categories and subcategories, with their counts (and percentages) of identified relevant papers (max. 77) and unique intents (max. 172). Note: Papers may include multiple unique intents and can therefore appear in multiple categories and subcategories.}
    \begin{tabular}{llccp{5cm}}
    \toprule
    \multirow{1}{*}{\textbf{Category}} & \multirow{1}{*}{\textbf{Subcategory}} & \multicolumn{2}{c}{\textbf{Number of}} & \multirow{1}{*}{\textbf{References}} \\
    & & Papers (\%) & Intents (\%) & \\
    \midrule
    \multirow{5}{*}{Motion} & \multirow{3}{*}{Robot Self-Actions} & \multirow{3}{*}{38 (49.35\%)} & \multirow{3}{*}{75 (43.60\%)} & \cite{Dragan.2013,Andersen.2016,Bolano.2021,Bolano.2018,Cakmak.2011,Capelli.2019,Chadalavada.2020,Chakraborti.2018,Che.2020,Cleaver.2021,Coovert.2014,Correa.2010,Domonkos.2020,Dragan.2015,Fischer.2016,Frederiksen.2019,Gielniak.2011,Gruenefeld.2020,Gu.2021,He.2020,Hetherington.2021,LeMasurier.2021,Matsumaru.2006b,Matsumaru.2007,Matsumaru.2005,Matsumaru.2006,Mikawa.2018,Rosen.2019,Ruffaldi.2016,Szafir.2015,Tsamis.2021,Walker.2018,Watanabe.2015,Wengefeld.2020,Zahray.2020,Zolotas.2019,Bodden.2018,Johannsen.2002} \\
    & \multirow{2}{*}{World Actions} & \multirow{2}{*}{15 (19.48\%)} & \multirow{2}{*}{18 (10.47\%)} & \cite{Andersen.2016,Aubert.2018,Chakraborti.2018,Chen.2011,Faria.2021,Faria.2017,Han.2020,Holladay.2014,Kebude.2018,Kirchner.2011,LeeChang.2018,Moon.2014,Newbury.2021,Palinko.2020,Psarakis.2022} \\
    \midrule
    \multirow{2}{*}{Attention} & Robot-Focused Attention & 6 (7.79\%) & 8 (4.65\%) & \cite{Aubert.2018,Bolano.2018,Cha.2016,Che.2018,Furuhashi.2015,Koay.2013} \\ 
    & World-Focused Attention & 4 (5.19\%) & 5 (2.91\%) & \cite{Levillain.2019,Mutlu.2009,Song.2018b,Staudte.2009} \\
    \midrule
    \multirow{4}{*}{State} & \multirow{3}{*}{Robot Self-Perception} & \multirow{3}{*}{23 (29.87\%)} & \multirow{3}{*}{27 (15.70\%)} & \cite{Andersen.2016,Baraka.2016,Cauchard.2016,Collins.2015,Correa.2010,Duncan.2018,Fletcher.2021,Gu.2021,Levillain.2019,Matsumaru.2006b,Matsumaru.2007,Novitzky.2012,Sharma.2013,Song.2018,Szafir.2014,Takayama.2011,Tang.2019,Walker.2018,Wengefeld.2020,Zolotas.2019,Johannsen.2002,Zhou.2017,Bacula.2020} \\ 
    & Robot World Perception & 8 (10.39\%) & 12 (6.98\%) & \cite{Andersen.2016,Chakraborti.2018,Coovert.2014,Correa.2010,Han.2020,Ruffaldi.2016,Wengefeld.2020,Zolotas.2019} \\ 
    \midrule
    \multirow{2}{*}{Instruction} & Robot-Centered Instructions & 10 (12.99\%) & 16 (9.30\%) & \cite{Baraka.2016,Cha.2016,Faria.2016,Furuhashi.2015,Glas.2007,Koay.2013,Levillain.2019,Mullen.2021,Song.2018c,Tang.2019} \\
    & World-Centered Instructions & 9 (11.69\%) & 11 (6.40\%) & \cite{Andersen.2016,Baraka.2016,Bolano.2021,Cakmak.2011,Chakraborti.2018,Chandan.2019,Moon.2014,Psarakis.2022,Wengefeld.2020} \\ 
    \bottomrule
    \end{tabular}
    \Description{Overview of different intent types, sorted by their categories and subcategories, with their counts (and percentages) of identified relevant papers (max. 77) and unique intents (max. 172). Each Category has two subcategories.}
    \label{tab:intenttypes}
\end{table*}

\paragraph{Motion}
\label{par:intentmodel:intent:definition:motion}
These intents are the main \emph{type of intent}.
\emph{Motion intent} is concerned with explicitly communicating future motions (i.e., actions that the robot will perform).
As our survey is focused on \emph{robot motion intent}, it encompasses more than 50\% of the identified unique intents in our corpus.
Most of the described intents deal with \emph{robot self-actions}, aiming to indicate future robot movement. 
Thereby, users may be able to improve the coordination of their actions in concert with the robot's behavior to avoid collisions and improve safety. 
For example,~\citeauthor{Chadalavada.2020} employed \ac{SAR} to communicate future movement direction as well as the specific path the robot will take, which helped \emph{bystanders} feel safe around a robotic forklift~\cite{Chadalavada.2020}.
\emph{World actions} are activities that manipulate the world around the robot. 
Again, this may help the \emph{bystander} to coordinate their activities, but it also helps the \emph{observer} to understand when to take over control from the robot.
\citeauthor{Psarakis.2022} applied this concept of \emph{world actions} in a VR simulation to visually augment the nearby objects that the robot planned to grasp~\cite{Psarakis.2022}.

\paragraph{Attention}
\label{par:intentmodel:intent:definition:attention}
Intents that communicate the need for attention are a supportive element. 
They precede a \emph{motion intent} to shift human attention toward the robot or process, especially when the humans' attention is not guaranteed (e.g., because they focus on their own tasks).
For example, \citeauthor{Bolano.2018} used acoustic feedback to alert the human and shift their attention toward the robot whenever it detected a possible collision~\cite{Bolano.2018}.
An example of \emph{robot-focused attention} was presented by~\citeauthor{Furuhashi.2015}, who designed an assistive robot based on the commercial Roomba device as a hearing dog that can notify deaf users of important events~\cite{Furuhashi.2015}. 
Here, the system uses physical touch to gain the human's attention by gently bumping into their body.
As an example of \emph{world-focused attention}, \citeauthor{Mutlu.2009} had a humanoid robot quickly look at an object of interest. They studied whether collaborators were able to understand the robot's gaze cues and correctly identify the object (among several others) that the robot had chosen as its object of interest~\cite{Mutlu.2009}).

\paragraph{State}
\label{par:intentmodel:intent:definition:state}
A robot communicating its state allows a \emph{human} to deduce potential future motions and identify conflicts before they occur.
For example, a \emph{robot} could collide with nearby objects due to errors in its sensor system. However, robot communication of the detected objects enables a \emph{human} to take over control and mitigate the issue.
For \emph{state} intents, we distinguish between \emph{robot self-perception}, meaning the state the \emph{robot} communicates about itself (e.g., simple text feedback presented on a display that indicates states such as \enquote{stop} or \enquote{moving}~\cite{Matsumaru.2007}), and \emph{robot world perception}, meaning the communication of the perceived state of the world (e.g., visually highlighting objects in the environment that the sensor system has successfully detected, allowing the user to predict and understand subsequent robot movement~\cite{Han.2020}).

\paragraph{Instruction}
\label{par:intentmodel:intent:definition:instruction}
In several papers, we identified \emph{instruction} intents that accompany robot motion.
For example, if a \emph{robot} is blocked by an obstacle, it can instruct a \emph{human} to remove the obstacle so it can continue its motion.
\emph{Instructions} can be \emph{robot-centered instructions} when they stand in relation to the robot itself (e.g., \citeauthor{Moon.2014} applied head gaze cues to communicate instructions to the user to complete the  handover of an object from the robot's gripper~\cite{Moon.2014}).
Or, in contrast, \emph{instructions} can be \emph{world-centered instructions} when they stand in relation to the world (e.g., a robot instructing a human to push a button on a wall to open an elevator so that it can continue its movement~\cite{Wengefeld.2020}).

\subsubsection{Relationship to Human}
\label{sub:intentmodel:intent:relationship}
Communicating a robot's intended motion to the human 
helps to improve the perception and understanding of the robot's behavior.
However, humans that are, for example, not involved in the robots' task -- perhaps because they are focusing on their own tasks (\emph{coworker}) or are just uninvolved in general (\emph{bystander}) -- often need an additional cue to be able to read \emph{robot motion intent}, which makes the intent type \emph{attention} necessary (e.g., by an acoustic prompt~\cite{Aubert.2018}).
\emph{State} intents enable a human to see not only the next \emph{motion} but also the internal state and planning, enabling them to understand actions ahead of time. Such intents also support \emph{observers} in their task of supervising the robot. 
Finally, collaboration means a constant shifting of who is in charge when humans and robots work together on a shared task. 
Therefore, \emph{motion}, \emph{state}, \emph{attention}, and \emph{instructions} are all necessary intents for providing a baseline for collaboration (\emph{collaborator}).

\subsection{Robot}
\label{sub:intentmodel:robot}
In our corpus, we identified three different \emph{kinds of robot}, which together form the \emph{robot} entity.

\subsubsection{Definition}
\label{sub:intentmodel:robot:definition}
We identified three main \emph{kinds of robots}: \emph{robotic arm}, \emph{humanoid}, and \emph{mobile robot}. 
These, in order, represent a spectrum of increasing mobility and flexibility based on the area of deployment, starting with stationary robots (still with many \ac{DoF}) and ending with robots that are inherently mobile (which also includes mobile arms with many \ac{DoF} on a platform). 
Based on different robots, researchers have investigated different intents with varying frequencies.
In the following, we illustrate each \emph{kind of robot} with examples from our literature corpus.

\paragraph{Robotic Arm}
\label{sub:intentmodel:robot:definition:arm}
\emph{Robotic arms} can be described as a chain of axis links. 
They are typically fixed to one place and can have a manipulator~\cite{Gautam.2017roboticArm}. 
Nowadays, they are the industry standard in production lines of factories~\cite{Braganca.2019} and work alongside humans in \ac{HRI} environments~\cite{Domonkos.2020}.
\emph{Robotic arms} are described in 13 papers and send 22 intents.

\paragraph{Humanoid}
\label{sub:intentmodel:robot:definition:humanoid}
\emph{Humanoids} have two robotic arms with manipulators, a torso, a head, eyes, and, often, basic facial expressions. 
Due to the two robotic arms, \emph{humanoids} have more \ac{DoF} than single robotic arms.
Still, \emph{humanoids} are often fixed to one place and lack mobility.
Nonetheless, they are an important part of \ac{HRI} when working with humans in a shared workspace~\cite{LeMasurier.2021,Rosen.2019}. 
In rare cases, they can move in space, imitating human movement. 
Here, anthropomorphic features of the robots -- such as gaze or certain gestures -- can decrease the time required to predict the robot's intent~\cite{Gielniak.2011}.
\emph{Humanoids} are described in 11 papers and send 21 intents.

\paragraph{Mobile Robot}
\label{sub:intentmodel:robot:definition:mobile}
With the addition of mobility comes increased flexibility. \emph{Mobile robots} can be deployed in the air, on the ground, or in water.
For this kind of robot, we have actively chosen to define them more broadly to include robots that appear only once in the corpus.
For \emph{mobile robots} (also referred to as drones), we distinguish between \emph{ground drones without a manipulator} that move between locations, \emph{ground drones with a manipulator} that can also manipulate the world, \emph{flying drones} that maneuver through the air, and \emph{water drones} that operate on water or underwater.
Communicating \emph{motion intent} helps \emph{ground drones without a manipulator} to, for example,  lead or follow a human to a specific place~\cite{Glas.2007}. It can help \emph{ground drones with a manipulator} to, for example, communicate which object they intend to pick up~\cite{Chakraborti.2018}.
\emph{Flying drones} or \emph{water drones}, on the other hand, can communicate their \emph{motion intent} by flying or driving in a pattern~\cite{Szafir.2014,Novitzky.2012}.
All kinds of drones can appear alone~\cite{Cleaver.2021} or as a swarm of drones~\cite{Capelli.2019}.
\emph{Mobile robots} are described in 53 papers and send 129 intents.

\subsubsection{Relationship to Intent}
\label{sub:intentmodel:robot:relationship}
As \emph{mobile robots} move around more freely, they frequently encounter human \emph{bystanders} who cross their paths.
Consequently, \emph{mobile robots} often have to first shift the \emph{human's} attention toward the robot's display, preparing them for the communication of the robot's intended \emph{motion}.
For example, a projection in front of the robot can catch the attention of a bystander while simultaneously informing about the direction of driving~\cite{Matsumaru.2006b}.
At the same time, \emph{mobile robots} need to communicate their \emph{state} and planning of actions ahead of time, either the inner state (e.g., what is the current mission status~\cite{Levillain.2019}) or the perceived world state (e.g., which objects are detected~\cite{Correa.2010}).
\emph{Humanoids} and \emph{robotic arms}, on the other hand, are often deployed in collaborative scenarios, teaming up with humans.
Here, robots need to communicate their intended \emph{motion} to coordinate their actions with a human collaborator (e.g., which items the robot intents to pick next from a shared bin~\cite{Aubert.2018} or when objects are to be handed over to the collaborator~\cite{Newbury.2021}). 

\subsection{Context}
\label{sub:intentmodel:context}
The \emph{context} describes the setting of the \ac{HRI} scenario.
While the location is an essential part of the context, there is more: for example, the social environment~\cite{Schmidt.1999context}.
Nonetheless, we consider the location helpful to define \ac{HRI} scenarios.
In our analysis, we found various types of locations, including \emph{workplace}, \emph{domestic}, and \emph{outdoor}.
In \emph{workplace} settings, the robot is frequently part of an assembly line or, more generically, a manufacturing process (e.g., collaborating with a human worker~\cite{Tang.2019}).
However, \emph{workplace} locations also include industrial settings, offices, or generic work rooms.
In total, 42 papers took place at a \emph{workplace} location.
In \emph{domestic} environments, robots support a task at home (e.g., by picking cups up off a kitchen table~\cite{Dragan.2015}).
Here, we found five relevant papers.
Finally, in two papers the robot could move freely outside (e.g., fulfilling a mission and communicating its status~\cite{Duncan.2018}).
Apart from these, 28 papers had no particular location specified. Instead, the authors of these papers investigate more generic scenarios of \emph{robot motion intent} (e.g., by stating that a robot moves between two locations but without fine details of these locations~\cite{Matsumaru.2007}).
For these scenarios, it is unclear which locations are most relevant.
\section{Analysis of Intent Information and Location}
\label{sec:indepthanalysis}
In addition to the different \emph{types of intent} discussed in the previous section, two other dimensions of intent emerged from the data: \emph{Intent information} (which refers to the data communicated by the \emph{robot}) and \emph{intent location} (which describes from where the intent is communicated to the \emph{human}). 
In this section, we define these dimensions, illustrate their application with examples, and present a summary of empirical findings concerning their usage.

\subsection{Intent Information}
\label{sub:indepthanalysis:information}
Based on our analysis of \emph{how} the intent is communicated as well as \emph{what} is communicated, we derived two main properties for categorizing \emph{intent information}: \emph{spatial} and \emph{temporal}.

\subsubsection{Spatial Property}
\label{sub:indepthanalysis:information:spatial}
The primary approach to convey spatial information is to embed it directly into the environment, i.e., have it \textbf{registered in space}. 
We identified 105 matching intents. 
We can further classify such intents as conveying \emph{local} information (74 intents) or \emph{directional} information (31 intents). 
\emph{Local} information aims to precisely relate the information to the surrounding space by showing an exact position that naturally may contain orientation information as well.
\citeauthor{Han.2020}, as an example, convey \emph{local} information by using \ac{SAR} polygon visualizations to frame and highlight detected objects on a table, allowing a human observer to supervise the robot's intended movement and manipulation actions~\cite{Han.2020}. 
In contrast, \emph{directional} information aims to communicate the explicit direction of movement (e.g., an arrow pointing in the direction of movement~\cite{Chadalavada.2020} or toward an object or person of interest~\cite{Holladay.2014}).

Information that is \textbf{unregistered in space}, however, employs an abstract encoding of the spatial property. In total, we identified 67 matching intents. This category includes the following \emph{types of intent}: \emph{Description}, \emph{symbol}, and \emph{signal}. \emph{Description} (11 intents) applies to scenarios in which textual or verbal information is used (e.g., the robot informs the human verbally before initiating a movement to perform a touch~\cite{Chen.2011}). \emph{Symbol} (25 intents) applies to cases in which a symbolic representation is used to form the intent communication (e.g., a mobile robot that nods its head to request a human follow before moving toward its destination~\cite{Faria.2016}). \emph{Signal} (31 intents) applies when components are turned on/off to indicate a change (e.g., an acoustic prompt is turned on to gain attention for the upcoming communication of \emph{motion intent}~\cite{Aubert.2018}). 
Mini maps provide an abstract but geographical encoding that includes the relationships among different objects in the environment \cite{Chandan.2019,Walker.2018,Zolotas.2019}.

\textbf{Empirical Implications.} 
While information \emph{registered in space} provides a direct link between real-world objects and the displayed information, information \emph{unregistered in space} lacks this connection and requires an additional mental step to create this link.
Consequently, information \emph{unregistered in space} may be less intuitive, and thus researchers have explored different combinations of information to mitigate that.
\citeauthor{Andersen.2016} as well as~\citeauthor{Wengefeld.2020} showed that combining multiple types of intent information that are \emph{unregistered in space} (e.g., text \emph{description} and \emph{symbol} icons) helps to effectively communicate \emph{motion intent} to the user~\cite{Andersen.2016,Wengefeld.2020}.
On the other hand, \citeauthor{Staudte.2009} found that combining both categories (\emph{registered \& unregistered}), which in their case involved a robot gazing at a specific object while a verbal description of the object played, leads to successful perception and understanding by the user~\cite{Staudte.2009}.
Similarly, \citeauthor{Bolano.2018} later showed that a verbal description of the target can be combined with visual feedback of the motion endpoint to achieve the same improvement~\cite{Bolano.2018}.

\subsubsection{Temporal Property}
\label{sub:indepthanalysis:information:temporal}
The temporal property of \emph{intent information} is about the distinction between having a \emph{discrete} or \emph{continuous} information flow.
\textbf{Discrete} information has a fixed, distinct appearance in time and is beneficial for communicating \emph{robot motion intent} because it enables the human to detect a change (i.e., the information appears) and it signals at which point the information loses its relevance (i.e., it disappears).
For example, \citeauthor{Aubert.2018} equip their humanoid robot with a display that shows the number of the next bin it will approach, thereby allowing a human to avoid conflict with the robot~\cite{Aubert.2018}. 
Overall, we identified 89 intents that communicate \emph{discrete} information. 
\textbf{Continuous} information, as has been provided in 83 intents, is available throughout the whole task or over several task phases (i.e., it is visible independent of its relevance to the current task).
It enables the human to observe the robot, compare it with the world, and evaluate the correct task execution.
\citeauthor{Tsamis.2021}, for example, implemented \ac{AR} visualizations for a \ac{HMD} to continuously communicate the intended movement space of a robotic arm by placing a semitransparent red sphere around the robotic arm~\cite{Tsamis.2021}.

\textbf{Empirical Implications.}
\citeauthor{Faria.2016} showed that both \emph{discrete} and \emph{continuous} information are effective for communicating a \emph{follow me} intent with spherical robots~\cite{Faria.2016}.
\citeauthor{Koay.2013} also evaluated both temporal properties using a robot dog that guides people living with hearing loss. However, they found that a motion-based approach (\emph{continuous}), in which the robot's head movements request users to follow, is more successful than using a flashing \ac{LED} stripe (\emph{discrete}). They attribute this to the fact that head movements are more straightforward to interpret~\cite{Koay.2013}.
The findings of \citeauthor{Aubert.2018} suggest that combining \emph{discrete} and \emph{continuous} information is the most effective method. They showed that the combination of a motion-based approach (\emph{continuous}) and a display approach (\emph{discrete}) to communicate the robot motion end-point outperformed both uni-modal intent communication conditions~\cite{Aubert.2018}.

\subsubsection{Cross Relations} 
\label{sub:indepthanalysis:information:crossrelation}
Inherently, the information of every intent has \emph{spatial} and \emph{temporal} properties. In the following, we describe the relationships between these properties of intent information.

For \emph{unregistered in space}, the temporal property is almost evenly distributed between \emph{discrete} and \emph{continuous} information. Here, \emph{signal} is an exception, as \emph{discrete} (23 intents; e.g., having flashing lights attached to a mobile robot to indicate a discrete change of movement direction, similar to a car~\cite{Hetherington.2021}) is used more often than \emph{continuous} (eight intents; e.g., an \ac{LED} stripe attached to the robot to continuously communicate the remaining distance to the target position through a color-coded progress bar~\cite{Baraka.2016}). 
\emph{Signals} are primarily used to communicate sudden changes. Accordingly, such \emph{discrete} events are naturally communicated as \emph{discrete} intent information.

For \emph{registered in space}, we see an uneven distribution for both subcategories.
Intent information classified as \emph{local} is mostly communicated  as \emph{continuous} information (50 intents; e.g., using \ac{SAR} to continuously highlight an area in a workplace where the robot will be active during its movements and action~\cite{Andersen.2016}) instead of \emph{discrete} (24 intents; e.g., using \ac{SAR} to highlight a button on a wall that must be pushed by a human for the robot to continue its movement~\cite{Wengefeld.2020}).
We think that robot \emph{motion} likely relates to a continuous event because it is meant to happen over time and takes place continuously.
Intent information classified as \emph{directional} is mostly communicated as \emph{discrete} information (23 intents; e.g., a display is attached to the top of a mobile robot, communicating the intended movement direction with an arrow~\cite{Matsumaru.2007}) and only seldom as \emph{continuous} (8 intents; e.g., a drone is visualized as an eye in \ac{AR}, constantly looking in the direction of movement~\cite{Walker.2018}).
The reason is that \emph{directions} are primarily used to communicate an updated movement direction to the human; therefore, it makes sense that they are most often given as \emph{discrete} information.

\subsection{Intent Location}
\label{sub:indepthanalysis:location}
Various technologies can enable the communication of \emph{robot motion intent}. 
We found that, in particular, the placement of these technologies (\emph{on-robot}, \emph{on-world}, and \emph{on-human}) can help to classify the different approaches in the literature, as there is often a relationship between the placement and specific types of technology. 

\emph{On-Robot} can be further divided into \emph{robot-only} technology or additional \emph{robot-attached} devices.
We identified 114 intents communicated through \emph{on-robot} technology.
As an example for the subcategory \emph{robot-only}, \citeauthor{Moon.2014} utilize the head orientation of the robot, mimicking a gaze cue, to communicate mid-air locations for its intended movement as an instruction to the user~\cite{Moon.2014}.
Nearly half of all categorized intents that utilize \emph{on-robot} technology fall into that subcategory, which is of particular interest because it limits the need for additional technology and often involves imitation of human-to-human behavior. 
The \emph{robot-attached} subcategory requires some additional hardware to be mounted to the robot (e.g., \ac{SAR}, \ac{LED}, or displays). 
For example,~\citeauthor{Wengefeld.2020} attach a laser projection system to the robot and thereby communicate various types of intents, including \emph{state}, \emph{motion}, and \emph{instruction}~\cite{Wengefeld.2020}.

\emph{On-World} has received relatively little attention in the literature. 
It includes, for example, small displays attached to the workspace at object bins~\cite{Aubert.2018}, or a desktop display (to visualize \emph{motion intent}) with speakers (to gain \emph{attention}) next to the robot's workspace~\cite{Bolano.2018}.
While the inability to change the environment may be less desirable from a generalizability perspective, for some technology, it adds significant benefits. In particular, \ac{SAR} would be easier to realize with a fixed projector position \emph{on-world} and it would allow for larger projection areas. We identified eight different intents \emph{on-world}.

\emph{On-Human} includes \emph{head-attached} technologies, which primarily refers to \ac{HMD} devices, which allow more complex visualizations. \citeauthor{Gruenefeld.2020}, for example, experimented with different spatial visualizations, such as visualizing the intended movement path, previewing future locations of the robot arm, or visualizing the activity area as a whole~\cite{Gruenefeld.2020}.
In addition, some approaches rely on \emph{hand-held} technologies.
\citeauthor{Correa.2010}, for example, used a tablet device displaying various types of information (map, live view, next steps) to support oversight and communicate \emph{motion intent}~\cite{Correa.2010}. 
We identified 50 intents \emph{on-human}.

\textbf{Empirical Implications.} 
For the \emph{intent location}, it is generally better to output information closer to the target.
For example, \citeauthor{LeMasurier.2021} compared several motion-based and light-based approaches for \emph{humanoids} to communicate an intended start of movement at an assembly workplace.
They saw that an \ac{LED} bracelet located closest to the workspace was the most noticeable and least confusing~\cite{LeMasurier.2021}. Furthermore, researchers found evidence that humans may prioritize \emph{on-human} technology over \emph{on-robot} technology. For example, \citeauthor{Che.2020} were able to show that the use of a vibrotactile bracelet worn by the user led to a better expression of the robot's \emph{motion intent}, reduced users' effort, and increased users' trust in the robot during a collision-avoidance movement when compared to a solely robot-based approach using \emph{legible motion}~\cite{Che.2020}. Finally, combining multiple output technologies can further increase performance. For example, \citeauthor{Mullen.2021} investigated a  multi-modal approach for communicating robot interference in a sorting scenario that combined an \ac{AR}-\ac{HMD} visualization and active feedback via a vibrotactile bracelet. They found that combining both feedback types outperformed the single modality baselines. It allowed the human to more efficiently teach the robot and decreased the required interaction time~\cite{Mullen.2021}.

\subsection{\change{Relation} between Location and Information} 
\label{sub:indepthanalysis:relationship}
In the following, we provide insights into the relationship between \emph{intent location} and \emph{intent information} (cf.~\autoref{tab:context:intentInformationToIntentTechnology}).

\begin{table*}
    \centering
    \caption{Overview of intents with different properties of \emph{intent information} (by rows) in combination with \emph{intent location} (by columns) -- up to three example references are listed for each category. Please note that each intent has a spatial and a temporal property.}
    \label{tab:context:intentInformationToIntentTechnology}
    \begin{tabular}{llccccc}
    \toprule
    \multirow{1}{*}{\textbf{Category}} & \multirow{1}{*}{\textbf{Subcategory}} & \multicolumn{2}{c}{\textbf{On-Human}} & \multirow{1}{*}{\textbf{On-World}} & \multicolumn{2}{c}{\textbf{On-Robot}} \\
    & & Head-Attached & Hand-Held & & Robot-Only & Robot-Attached \\
    \midrule
    \multirow{2}{1.55cm}{(Spatial) Registered} & Local &       
    \textbf{35}~\cite{Gruenefeld.2020,Rosen.2019,Walker.2018} & \textbf{3}~\cite{Correa.2010,Watanabe.2015} & \textbf{4}~\cite{Aubert.2018,Bolano.2018,Cleaver.2021} & \textbf{22}~\cite{Dragan.2015,Bodden.2018,Cakmak.2011} & \textbf{10}~\cite{Coovert.2014,Hetherington.2021,Wengefeld.2020} \\
    & Directional &                                            
    \textbf{3}~\cite{Gu.2021,Ruffaldi.2016,Walker.2018} & \textbf{0} & \textbf{0} & \textbf{14}~\cite{Holladay.2014,Mikawa.2018,Moon.2014} & \textbf{14}~\cite{Chadalavada.2020,Hetherington.2021,Matsumaru.2007} \\
    \midrule
    \multirow{3}{1.55cm}{(Spatial) Unregistered} & Description &   
    \textbf{0} & \textbf{1}~\cite{Correa.2010} & \textbf{1}~\cite{Bolano.2018} & \textbf{0}  & \textbf{9}~\cite{Matsumaru.2006b,Staudte.2009,Wengefeld.2020}  \\
    & Symbol &                                               
    \textbf{5}~\cite{Walker.2018,Zolotas.2019} & \textbf{0} & \textbf{1}~\cite{Chandan.2019} & \textbf{14}~\cite{Glas.2007,Koay.2013,LeMasurier.2021} & \textbf{5}~\cite{Andersen.2016,Bacula.2020,Song.2018c}  \\
    & Signal &                                               
    \textbf{0} & \textbf{3}~\cite{Che.2020,Che.2018,Mullen.2021} & \textbf{2}~\cite{Aubert.2018,Bolano.2018} & \textbf{0}  & \textbf{26}~\cite{Domonkos.2020,Szafir.2015,Tang.2019} \\
    \midrule
    \textbf{Total} & & \textbf{43} (25.00\%) & \textbf{7} (4.07\%) & \textbf{8} (4.65\%) & \textbf{50} (29.07\%) & \textbf{64} (37.21\%) \\
    \midrule
    \multirow{2}{1.55cm}{(Temporal) Discrete} & \multicolumn{1}{c}{\multirow{2}{*}{}} &    
    \multirow{2}{*}{\textbf{15}~\cite{Gu.2021,Newbury.2021,Psarakis.2022}} & \multirow{2}{*}{\textbf{4}~\cite{Che.2018,Che.2020,Mullen.2021}} & \multirow{2}{*}{\textbf{5}~\cite{Aubert.2018,Bolano.2018}} & \multirow{2}{*}{\textbf{19}~\cite{LeMasurier.2021,Furuhashi.2015,Gielniak.2011}} & \multirow{2}{*}{\textbf{45}~\cite{Cha.2016,Faria.2016,Zahray.2020}} \\
    &&&&&& \\
     \midrule
    \multirow{2}{1.55cm}{(Temporal) Continuous} & \multicolumn{1}{c}{\multirow{2}{*}{}} &       
    \multirow{2}{*}{\textbf{28}~\cite{Chakraborti.2018,Tsamis.2021,Zolotas.2019}} & \multirow{2}{*}{\textbf{3}~\cite{Correa.2010,Watanabe.2015}} & \multirow{2}{*}{\textbf{3}~\cite{Bolano.2018,Chandan.2019,Cleaver.2021}} & \multirow{2}{*}{\textbf{31}~\cite{Dragan.2015,Capelli.2019,Cauchard.2016}} & \multirow{2}{*}{\textbf{19}~\cite{Collins.2015,Han.2020,Matsumaru.2005}} \\
    &&&&&& \\
    \bottomrule
    \end{tabular}
    \Description{Overview of intents with different properties of intent information (by rows) in combination with intent location (by columns) – up to three example references are listed for each category. The table has an upper part for the spatial property and a lower for the temporal property. In the middle between them, the total numbers are shown.}
\end{table*}

\subsubsection{Registered in Space} 
To communicate location information registered in space, most researchers rely on \emph{head-attached} technologies, such as \ac{AR}-\ac{HMD}s (\emph{on-human}).
For example, \citeauthor{Tsamis.2021} implemented \ac{AR} visualizations to communicate an intended movement trajectory of a robotic arm~\cite{Tsamis.2021}.
They placed small spheres along a defined path in 3D space from the robot's end-manipulator to a specific destination. 
They found that using their system improved task completion and robot idle times, with fewer interruptions to the overall workflow. 
In addition, users reported increased feelings of safety and trust toward the robot.
In contrast, \citeauthor{Correa.2010} proposed a tablet visualization that showed a live camera feed of the mobile robot highlighting recognized objects in its environment via a wireframe in the visualization~\cite{Correa.2010}.
In addition to intents displayed \emph{on-human}, robots are often used to convey information directly through specific movements or pointing (\emph{on-robot}). 
For example, \citeauthor{Holladay.2014} used a robotic arm and its end-effector to communicate a directional cue by pointing toward an object placed on a table~\cite{Holladay.2014}.
The resulting pointing configurations were reported to make it easier for novice users to infer the target object.
Another example for displaying information \emph{on-robot} is provided by \citeauthor{Hetherington.2021} 
They used \ac{SAR} to project an arrow in the intended movement direction of the mobile robot on the floor~\cite{Hetherington.2021}.
Their results show that projected arrows were more socially acceptable and more understandable than flashing lights.
Finally, information \emph{registered in space} can be outputted \emph{on-world}.
For example, \citeauthor{Cleaver.2021} used their web-based environment~\cite{Cleaver.2020heaven} to compare four different conditions of visualizing the intended movement trajectory of a mobile robot on a \emph{world}-located display~\cite{Cleaver.2021}. 
In contrast, \citeauthor{Aubert.2018} placed small displays on three bins and used bin numbers and progress bars to indicate from which bin the robot coworker would next withdraw an item. 
However, the display-based approach could not significantly reduce the number of physical conflicts~\cite{Aubert.2018}.

\subsubsection{Unregistered in Space.}
Interestingly, a relatively large number of \emph{symbol} information is communicated through the robot itself (\emph{on-robot}). 
Here, we found many approaches where the robot performs specific movement patterns that the human has to decode appropriately.
A symbolic approach is shown by \citeauthor{LeMasurier.2021}~\cite{LeMasurier.2021}. They slightly move the robot's manipulator to the left and right to communicate an intended movement start. 
This approach received relatively high ratings on several measures; however, the authors recommend that the addition of light signals near the workspace and the origin of motion (like an \ac{LED} bracelet) may provide a benefit to \ac{HRI} in shared spaces.  
\citeauthor{Song.2018c} provide an example of the type \emph{symbol} by using different static and dynamic light patterns on a \emph{robot-attached} colored \ac{LED} stripe to illustrate different \emph{states} of the robot~\cite{Song.2018c}.
Communication of \emph{signal} information is mainly achieved through robot-attached technology, such as \ac{LED} or audio speakers.
Wearable technologies can also show spatially \emph{unregistered} information (\emph{on-human}).
\citeauthor{Che.2020} propose a vibrotactile bracelet worn by the user to communicate an initiated collision-avoidance movement of a \emph{mobile robot}~\cite{Che.2020}. 
This approach led to a better expression of the robot's \emph{motion intent}, reduced users' effort, and increased users' trust in the robot.
Furthermore, \citeauthor{Walker.2018} implemented a radar-like mini-map in the corner of an \ac{AR} visualization to illustrate the relative position of the user to a drone~\cite{Walker.2018}. 
Although the radar provides the user with the means to rapidly locate the robot relative to their own position, some participants mentioned that they did not need to use the radar much because they always faced the drone.
Finally, unregistered information can also be presented \emph{on-world}.
\citeauthor{Bolano.2018} propose verbally describing the updated destination of the robot's end-manipulator via a speaker in addition to the screens placed in the shared workspace~\cite{Bolano.2018}. 
They found that users better understood the robot's intended motion, including when the robot had to reroute itself to avoid collision.

\subsubsection{Discrete.} 
\emph{Discrete} information is usually presented directly \emph{on-robot}.
As an example of \emph{robot-attached} technology, \citeauthor{Domonkos.2020} attached a colored \ac{LED} stripe to the base of a robotic arm to communicate the intended direction of movement to a human \emph{coworker}~\cite{Domonkos.2020}. In contrast, \citeauthor{Glas.2007} proposed a \emph{mobile robot} that performs head gestures to initiate either a follow-me or lead-me request to the human~\cite{Glas.2007}, relying on the robot itself as in \emph{robot-only}.
\citeauthor{Gu.2021} evaluated a visual feedback displayed through an \ac{AR}-\ac{HMD} (\emph{on-human}), indicating the planned movement direction of the robot via an arrow visualization~\cite{Gu.2021}. They found that the visualization improved perceived safety and task efficiency. Instead of relying on the visual modality, 
\citeauthor{Mullen.2021} proposed discrete feedback through a vibrotactile bracelet that is activated to communicate robot interference, triggering the human to move in order to allow the robot to continue its movement~\cite{Mullen.2021}.
Their findings show that vibrational feedback can reduce the time required to notice and respond to an intent.  
\citeauthor{Aubert.2018} equipped bins (from which items could be chosen) in the environment with speakers to emit \emph{discrete} auditory information \emph{on world}~\cite{Aubert.2018}.
They recommend not solely relying on auditory information, but using it in a multi-modal approach, which is further supported by \citeauthor{Bolano.2018}~\cite{Bolano.2018}.

\subsubsection{Continuous.} 
Like \emph{discrete} information, \emph{continuous} information is primarily displayed \emph{on-robot}.
\citeauthor{Matsumaru.2005} \emph{attached} an omnidirectional display \emph{on-robot}, projecting an eyeball-like visualization that effectively communicates the direction of movement to a human~\cite{Matsumaru.2005}.
In contrast, \citeauthor{Dragan.2015} propose performing legible motions with a robotic arm itself to communicate the next object it will grasp~\cite{Dragan.2015}, which they found enabled fluent collaboration.
As an example of communicating intents \emph{on-human}, \citeauthor{Walker.2018} display a symbolic representation of a focusing eye lens in an \ac{AR}-\ac{HMD}, encoding the relative distance to the next target~\cite{Walker.2018}. Their results show a significant improvement in users' understanding of \emph{robot motion intent}.
\citeauthor{Watanabe.2015} proposed presenting \emph{continuous} visual feedback via a tablet to inform a wheelchair passenger of a robot's intended motion path~\cite{Watanabe.2015}.
Lastly, \emph{continuous} information can be displayed \emph{on-world}.
\citeauthor{Chandan.2019} proposed a map visualization for a stationary tablet display that continuously shows the locations of three mobile robots and other objects of interest~\cite{Chandan.2019}. They found this approach significantly improved the participants' ability to observe and assist the robot. Similarly, albeit only studied in a web-based experiment, \citeauthor{Cleaver.2021} proposed a 3D visualization displayed on a 2D screen to continuously communicate the intended path of a mobile robot~\cite{Cleaver.2021}.
\section{Discussion and Future Research}
\label{sec:discussion}
In the following, we discuss key findings of our literature survey and formulate future research directions \change{as takeaway messages for the \ac{HCI} community}.
The organization of the section follows the three entities \emph{human}, \emph{intent}, and \emph{robot} from our intent communication model \change{and concludes with a discussion of the overall model}.

\paragraph{Human}
From the analyzed intents of our corpus, we derived four different \emph{roles of human} (\emph{collaborator}, \emph{observer}, \emph{coworker}, and \emph{bystander}). 
In our analysis, we found that the human role is strongly related to the overarching goals of communicating \emph{motion intent} -- a specific goal can be directly derived given a specific human role.
For example, if the \ac{HRI} scenario involves the human taking the role of an \emph{observer}, the \emph{motion intent} needs to help with fostering oversight. 
As a result, this indicates that practitioners and researchers should explicitly define the role and, thereby, the involved human stakeholders before settling on the robot or specific intents they may want to communicate. 
The human roles we found in a bottom-up process through our analysis align well with the previous work of \citeauthor{Onnasch.2020rolesOfHuman}~\cite{Onnasch.2020rolesOfHuman}.
In contrast to \citeauthor{Onnasch.2020rolesOfHuman}, the role of the \emph{operator} did not show up in our analysis. 
We suggest this is because robots are not manually operated by humans in our corpus, as this would not require the robot to communicate any intent~\cite{grudin2017tool}.

\textbf{Future Research}: Our analysis showed that nearly all papers a) investigate individual human roles, e.g., they (often implicitly) pick one and focus on that, and b) design and study only for a 1:1 relationship between human and robot. 
The only exceptions to this are \citeauthor{Faria.2017}, \citeauthor{Kirchner.2011}, and \citeauthor{Palinko.2020}, who investigate the legibility of robot movement for a group of humans~\cite{Faria.2017} or explore the use of gaze cues to allow the robot to choose their human collaboration partner from a group of humans~\cite{Kirchner.2011, Palinko.2020}. 
This limited involvement of multi-user groups is, of course, to be expected in an emerging field that first needs to establish certain ground truths. 
Involving multiple persons or even multiple robots and persons complicates \ac{HRI} tremendously, yet we think this is the subsequent step research must take. 
In particular, it would be interesting to reflect on the suitability of specific technologies (e.g., \ac{SAR} will likely be better suited to satisfy multi-user scenarios compared to \ac{HMD} technology).

\paragraph{Intent Types}
Through our scoping review of \emph{robot motion intent}, we observed that communication of motion often requires additional intents that serve as pre- or post-cursors to the communicated \emph{motion intent}.
Furthermore, we found that robot motion can also be indirectly communicated: For example, by communicating only the robot's state (e.g.,~\cite{Baraka.2016}) or by instructing a human to open a door so the robot can continue on its path (e.g.,~\cite{Watanabe.2015}). 
These various \emph{types of intent} demonstrate the different facets of \emph{robot motion intent}, which represent both actual intended movement trajectories and related communication.
We see that as a key finding, distinguishing our work from previous research that focuses primarily on the communication of \emph{motion intent}~\cite{Rosen.2019,Suzuki.2022.AR-HRC-Survey,Walker.2018}. 
With our survey, we are confident that other researchers will start to adopt a more holistic and precise use of the term \emph{robot motion intent} and, for example, start highlighting the need for related intents, as we found in our analysis.

\textbf{Future Research}: Researchers should investigate how the different \emph{types of intent} may best be combined to achieve specific intent communication goals. 
Currently, there is little empirical knowledge about, for example, when and to what extent a robot may need to first communicate \emph{attention} before effectively being able to communicate \emph{motion intent}.
Further research should also challenge our classification of \emph{types of intent} and potentially extend them.

\paragraph{Intent Information and Location}
We derived two main properties that categorize our identified \emph{intent information} related to space: \emph{registered in space} (61.05\%) and \emph{unregistered in space} (38.95\%).
This almost-even distribution reveals that a lot of relevant research not only focuses on information that aims to convey \emph{local} or \emph{directional} information (e.g., a resulting trajectory~\cite{Cleaver.2021}), but also on more abstract representations, namely \emph{description}, \emph{symbol}, and \emph{signal}. 
These are often much less complex and indicate that \emph{robot motion intent} can be communicated without visual 3D representations of future movement.
This shows that there are viable alternatives to wearing special \emph{on-body} technology, resulting in fewer system costs and a decreased setup time. 
An alternative can be the \emph{intent location} \emph{on-robot}. 
In previous work, researchers have refined robots with anthropomorphic elements -- such as eye-like features or certain movement gestures -- to communicate motion intent.
Our literature review identified 15 such instances, specifically applying eye- or head-gaze (e.g., looking at an object to indicate a handover between human and robot~\cite{Moon.2014}).
While anthropomorphic elements may not be as precise as digital representations through technology means (e.g., visualizations in \ac{AR}), they share the same baselines as in \ac{HHC}.
The general assumption is that, in turn, they can be easily understood by users and can mostly be integrated into the actual \ac{HRI}. 
A possible combination with a verbal description provides a multi-modal output to the user, resulting in faster recognition of the specific object~\cite{Staudte.2009}.

\textbf{Future Research}: While previous research has explored combinations of spatially registered and unregistered information~\cite{Staudte.2009}, we are unaware of research that has contrasted their effectiveness.
Therefore, current design decisions may be based more on the availability of particular technology and less on the intended outcome. 
Future research should explore this further so that practitioners can more accurately judge the potential trade-offs between simple or complex information and related technology use.
Regarding the use of anthropomorphic features, the integration of such communication cues has been explored regarding their legibility and effectiveness in communicating \emph{robot motion intent}. However, their implicit consequences (e.g., causing the human to ascribe human-like behavior to the robot) may still need to be fully explored. The means and cues of communication have significant consequences for the trust relationship between humans and robots~\cite{Hamacher.2016}.

\paragraph{Robot}
When looking at the three \emph{kinds of robots} and their usage in research, we can see that the physical properties of a robot have a large impact on communication means: In particular, the \emph{on-robot} location for intent communication.
Some robots come with pre-installed displays, while others have anthropomorphic features built in. 
\emph{Flying drones}, on the contrary, require some kind of remote communication tool (often in the form of \ac{HMD}s) to communicate over a larger distance. 
Robots are also an area of much technical experimentation, i.e., many researchers are building or customizing their own robots.
For example, one may add anthropomorphic features to a robotic arm.  As a result, researchers tend to use these built-in or customized features to communicate intent. They may often have only a particular kind of robot available; thus, they are limited to a certain way of communicating \emph{robot motion intent}. 
Of course, this limits the generalizability of current findings, as each robot conveys unique features that can impact \ac{HRI}.

\textbf{Future Research}: These findings show that many research endeavors explore only certain \emph{kinds of robots}. 
A more systematic approach is called for to investigate the various kinds of robots and their impacts on communicating \emph{robot motion intent}.
We also found that more and more research applies simulation environments in \ac{VR} to explore \ac{HRI}.
Nevertheless, we need more studies to validate such findings and provide a broader foundation for their generalizability.

\paragraph{Context} 
Compared with previous research in \change{\ac{AVs}~\cite{Colley2021.AV.3AC,Currano2021.AV.3AC} and} \ac{eHMIs}~\cite{Dey.2020.AutomotiveSurvey}, we can identify several similarities, despite the substantial differences in the context of use and robot technology. \change{\citeauthor{Colley2021.AV.3AC} found that visualizing internal information processed by an \ac{AV} could calibrate  trust by enabling the perception of the vehicle’s detection capabilities (and its failures) while only inducing a low cognitive load~\cite{Colley2021.AV.3AC}. \citeauthor{Currano2021.AV.3AC} explored the interaction between complexity of head up displays, driving style, and situation awareness~\cite{Currano2021.AV.3AC}.
In the area of \ac{eHMIs},} researchers have been able to distinguish between different \emph{natures of message} (e.g., danger and safety zones)~\cite{Dey.2020.AutomotiveSurvey}.
These correspond to our identified \emph{types of intent}, highlighting different meanings for the user for the provided intent. 
In the context of \ac{AVs}, the information used to formulate the actual intent is primarily unregistered in space. 
It uses text, symbols, and audio prompts. 
The intent primarily describes the vehicle's state (e.g., automated/manual, cruising, yielding) or advice/instructions to the pedestrian (e.g., to allow safe road crossing).
The large differences between the fields of research result primarily from the standardizations in automotive research, such as roads, road signs, markings, and restrictions.
Nevertheless, there are potential overlaps.

\textbf{Future Research}: The two fields have, from our perspective, not yet shared many cross-activities among researchers, which could lead, for example, to transferring those \emph{motion intent} techniques that have shown to be effective in one field to the other. 
We could imagine that future research could benefit both sides if a more holistic perspective is applied. 
In particular, the research for \ac{eHMIs} in \ac{AVs} could benefit from more exploratory technological approaches in \ac{HRI}, such as making use of \ac{AR}-\ac{HMD}s and applying more advanced visualization to communicate \emph{motion intent}.
While this may not be relevant for the near future, as such devices are not yet consumer-ready, this may change over the coming years. 

\change{
\paragraph{The Model}
The overall model is an abstract characterization of the current literature on \emph{robot motion intent}. It may be seen as a summary of the current understanding of the design space for robot intent communication, where it illustrates all components and highlights their interconnection.
Thereby, future researchers and practitioners should benefit from the model by using it as a guidance and checklist throughout the design phase of such Human-Robot scenarios; i.e., being guided to carefully think and decide upon different types of intents or whether intent information should be encoded spatially or temporally. In addition, the model can help to unify the language of \emph{robot motion intent} and thereby support researchers and practitioners to find related work as well as help to identify research gaps.

\textbf{Future Research:} We invite researchers to actively challenge the model and thereby helping to develop the field even further. They should scrutinize whether the design space is sufficiently classified or how it can and needs to be extended to cover future work. As our model was derived from the analysis of our literature corpus, it is fitted to the gathered research. Nonetheless, one can utilize novel research contributions that will be published in the future to revisit and evaluate the model (i.e., to investigate if novel contributions can still be described by our model). Moreover, we imagine that a more thorough discussion in the context of eHMIs may benefit the model as well as incorporating other lines of research that are concerned with communicating intent, such as \citeauthor{Sodhi2012} or \citeauthor{Mueller2020}~\cite{Sodhi2012,Mueller2020}.}
\section{Conclusion}
\label{sec:summary-and-conclusion}
This paper provides two main contributions: 1) a survey contribution that includes an analysis and classification of previous literature as well as future research directions, and 2) a theoretical contribution that introduces an intent communication model and describes the relationships of its entities, dimensions, and underlying properties.
In particular, our work highlights that \emph{robot motion intent} requires a broader perspective on robot intent and that it includes intent types that may seem, at first glance, unrelated to motion. 
However, in our analysis, we found that \emph{attention}, \emph{state}, and \emph{instruction} are important and often necessary pre- or post-cursors to communicate explicit \emph{motion intent}. We also found that only a few papers explicitly discuss or present the type of intent they aim to communicate and they also lack clear descriptions of intent information or location. 
Our work aims to help researchers in the future to better align their work with the suggested dimensions, making it easier to assess and compare different studies.
Therefore, we aim to provide a foundation for a unified language regarding \emph{robot intent}, even beyond motion. 
From a practical perspective, the classification of the existing research literature along our \emph{intent communication model} helps researchers and practitioners alike to understand the design space for communicating \emph{robot motion intent}.
As it is an emerging field, much work has focused on finding novel approaches and solutions to communicate \emph{robot motion intent} in one way or another. 
We have identified multiple areas of need for future research directions. However, we would like to emphasize once more that, above all, the field needs more systematic analysis and comparison of different approaches to improve understanding of the influences of different intent dimensions and properties. 
We believe that the presented intent communication model provides an empirically deducted foundation to inspire and guide such work.



\bibliographystyle{ACM-Reference-Format}
\bibliography{bibliography}


\end{document}